\documentclass[aps,twocolumn,superscriptaddress,amsmath]{revtex4-2}

\usepackage{graphicx}
\usepackage[utf8]{inputenc}
\usepackage[colorlinks,allcolors=blue]{hyperref}
\usepackage{braket}
\usepackage{placeins}
\usepackage{amsmath}
\DeclareMathOperator{\Tr}{Tr}
\usepackage{xcolor}
\usepackage{comment}
\begin{document}

\newcommand{\mytitle}{Universal stability of coherently diffusive 1D systems with respect to decoherence.}

\title{\mytitle{}}

\author{F. S. Lozano-Negro}
\affiliation{Instituto de F\'isica Enrique Gaviola (CONICET-UNC) and Facultad de Matem\'atica, Astronom\'ia, F\'isica y Computaci\'on, Universidad Nacional de C\'ordoba, 5000, C\'ordoba, Argentina}

\author{E. Alvarez Navarro}
\affiliation{Benem\'erita Universidad Aut\'onoma de Puebla, Apartado Postal J-48, Instituto de F\'isica, 72570, Mexico}

\author{N. C. Ch\'avez}
\affiliation{Dipartimento di Matematica e Fisica and Interdisciplinary Laboratories for Advanced Materials Physics, Universit\`a Cattolica, via della Garzetta 48, 25133 Brescia, Italy}

\author{F. Mattiotti}
\affiliation{University of Strasbourg and CNRS, CESQ and ISIS (UMR 7006), aQCess, 67000 Strasbourg, France}

\author{F. Borgonovi}
\affiliation{Dipartimento di Matematica e Fisica and Interdisciplinary Laboratories for Advanced Materials Physics, Universit\`a Cattolica, via della Garzetta 48, 25133 Brescia, Italy}
\affiliation{Istituto Nazionale di Fisica Nucleare, Sezione di Milano, via Celoria 16, I-20133, Milano, Italy}

\author{H.~M.~Pastawski}
\affiliation{Instituto de F\'isica Enrique Gaviola (CONICET-UNC) and Facultad de Matem\'atica, Astronom\'ia, F\'isica y Computaci\'on, Universidad Nacional de C\'ordoba, 5000, C\'ordoba, Argentina}

\author{G. L. Celardo}
\affiliation{Department of Physics and Astronomy, CSDC and INFN, Florence Section, University of Florence, Italy}

\begin{abstract}
Static disorder in a 3D crystal degrades the ideal ballistic dynamics until it produces a localized regime. This Metal-Insulator Transition is often preceded by coherent diffusion.
By studying three paradigmatic 1D models, namely the Harper-Hofstadter-Aubry-André and Fibonacci tight-binding chains, along with the power-banded random matrix model, we show that whenever coherent diffusion is present, transport is exceptionally stable against decoherent noise. This is completely at odds with what happens for coherently ballistic and localized dynamics, where the diffusion coefficient strongly depends on the environmental decoherence. A universal dependence of the diffusion coefficient on the decoherence strength is analytically derived: the diffusion coefficient remains almost decoherence-independent until the coherence time becomes comparable with the mean elastic scattering time. 
 Thus, systems with a quantum diffusive regime could be used to design robust quantum wires. Moreover our results might shed new light on the functionality of many biological systems, which often operate at the border between the ballistic and localized regimes. 
\end{abstract}

\maketitle

\section{Introduction} 

The understanding and control of quantum transport in presence of environmental noise is crucial in many areas such as cold atoms~\cite{3akkermans2008photon}, mesoscopic systems~\cite{4beenakker1997random}, and quantum biology~\cite{vattay2015Bio,2celardo2012superradiance}. Its better understanding would allow us to design more efficient sunlight harvesting systems~\cite{2spano1991cooperative,2mohseni2008environment,biolaser2021}, devices that transfer charge or energy with minimal dissipation~\cite{1schachenmayer2015cavity,chavez2021disorder} and bio-mimetic photon sensors~\cite{mohan2022}, as well as to explain the functionality of many biological aggregates~\cite{Naam12,gulli2019macroscopic,cao2016quantum,ACS2010}. 

\begin{figure}[h!]
\centering
\includegraphics[width=1\columnwidth]{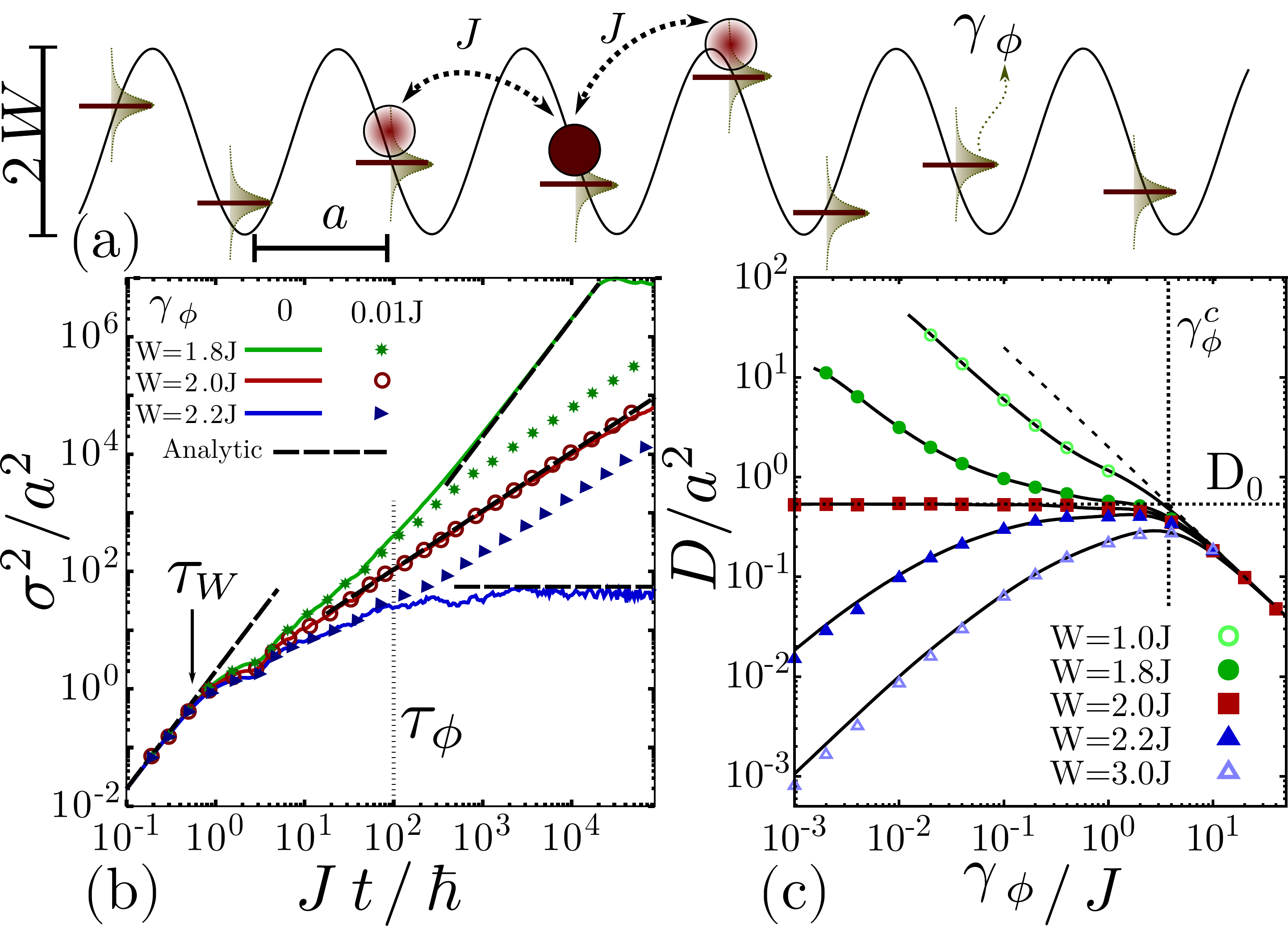} 
\caption{\textbf{(a)} HHAA model: horizontal lines are site energies, Lorentzian uncertainties indicate decoherence, $J$ is the hopping amplitude. \textbf{(b)} Excitation spreading for the HHAA model with $N=10000$ for different $W$. Coherent dynamics are shown with solid curves, while symbols show decoherent ones with a fixed value of $\gamma_\phi=0.01J$. Both symbols and curves share the same color (gray-tone) to denote a given $W$. The vertical arrow shows the \textit{mean elastic scattering time} \(\tau_W\), Eq.~\eqref{LET:eq:tw}. Black dashed lines are the analytical estimates for $\gamma_\phi=0$, see text. The vertical dotted line shows the decoherence time $\tau_\phi=\hbar/\gamma_\phi$. \textbf{(c)} Scaled diffusion coefficient $D/a^2$ vs. the decoherence strength $\gamma_\phi/J$ for different $W$. Solid black curves result from Eq.~\eqref{Eq:DSigama0-gen}. Different regimes are extended (green circles), critical (red squares), and localized (blue triangles). The horizontal black dotted line is the coherent theoretical estimate $D_0/a^2$; the black dashed line is the asymptotic $D/a^2\simeq 2J^2/\hbar\gamma_\phi$, and the vertical dotted line is the characteristic decoherence $\gamma^c_\phi=2\hbar/\tau_W$. Numerical data obtained by the QD method (symbols) for $N=1000$. In all panels: $q=(\sqrt{5}-1)/2$, $J=1$ and $\hbar=1$.}%
\label{fig:LET:intro}
\end{figure}

It was P. W. Anderson~\cite{AAnLR79} who realized that elastic scattering from a random disorder exceeding a critical value induces the localization of quantum excitations, and a \textit{metal-insulator transition} (MIT). While in 3D this critical disorder is finite, in 1D any amount of disorder is enough to localize.
Two decades later it was realized that correlated disorder and long-range hopping could allow a MIT even in 1D~\cite{AuAd80,DuWPh90,izrailev1999cor,mirlin_transition_1996}.

The different roles of the environment were considered by R. Landauer~\cite{La70}, N. Mott~\cite{Mott68}, and H. Haken~\cite{Haken58}. Specifically, Landauer noticed that an actual finite system exchanges particles with external reservoirs through the current and voltage probes, a notion that M. Büttiker used to describe environmental decoherence and thermalization~\cite{Bue86,DAPa90}. Both, Haken and Mott sought to address the role of a thermal bath. A very simple but widely used model for the environmental bath is the Haken-Strobl model, which describes uncorrelated dynamical fluctuations in the site energies.

Later on, Mott predicted a \textit{variable-range-hopping} regime in which energy exchange among phonons and Anderson's localized states would favor conductivity before decoherence freezes the dynamics~\cite{PaUs98}. Thus, in the localized regime, the 1D conductivity reaches a maximum~\cite{ThKi81,DAPa90,2mohseni2008environment,caruso2009} when the energy uncertainty associated with elastic scattering and that resulting from the coupling with the environment (i.e. decoherence processes)~\cite{moix2013coherent,chuang2016quantum} become comparable. In contrast, the ballistic dynamics of a perfect crystal is always degraded by the thermally induced decoherent scattering processes \cite{madhukar1977exact}.
{\it A much less studied subject is how decoherent noise affects transport 
around the MIT and, more generally, in presence of a quantum diffusive-like dynamics.}

Recent works on excitonic transport in large bio-molecules such as photosynthetic antenna complexes seek to explain the puzzling great efficiency of many natural~\cite{molina2016superradiance, caruso2009,2mohseni2008environment,LindsayVattay17} and biomimetic systems. 
In this context, S. Kauffman~\cite{Kau19} proposed the intriguing \textit{poised realm} hypothesis that, in biological systems, excitation transport occurs at the \textit{edge of chaos}. This led Vattay and coll.~\cite{vattay2014quantum} to propose that 1D systems near the MIT are optimal for transport because environmental decoherence does not affect the system as strongly as it does in the extended regime while it ensures delocalization needed for transport. 

This hypothesis seems at odds with an early theoretical analysis~\cite{Pa91} indicating that it is the intrinsic diffusive dynamics of some 1D systems what yields a particular stability of transport towards decoherence. 
With the purpose to settle this conflict, we study a few paradigmatic models that afford coherent diffusion. We first analyze the Harper-Hofstadter-Aubry-André (HHAA) model~\cite{AuAd80}, see Fig.~\ref{fig:LET:intro}a, which has been brought to the spotlight~\cite{torres2017dynamical,xu2019butterfly,lozano2021ergodicity} thanks to various experimental implementations~\cite{Nroati2008anderson,Sc+Bl15}. In absence of external noise, we found, both numerically and analytically, that only at the MIT the second moment of an initially localized excitation can be described by a diffusion coefficient $D$. There, as long as the decoherence strength remains below a characteristic value $\gamma_\phi^c$, see Fig.~\ref{fig:LET:intro}c, $D$ is very weakly dependent on the decoherent noise. On the other side, transport properties both in the extended and localized regimes are strongly affected by decoherence.
We also found that, at long times, $D$ determines the current and the system reversibility assessed by the \textit{Loschmidt echo} (LE) decay. Thus, at the MIT, both magnitudes are almost independent of the decoherent noise strength (see Appendix~\ref{Sup:Current}~\&~\ref{SM:purity}).
However, these findings do not settle the question of whether it is the diffusive quantum dynamics what brings stability towards decoherence or if this stability is inherent to the critical point.
For this reason we also studied the Fibonacci chain~\cite{Fibonacci1983} and the \textit{power-banded random matrices} (PBRM)~\cite{mirlin_transition_1996}, where a diffusive-like regime exists in some parameter range independently of their criticality. Our results show that, whenever a system is in a quantum coherent diffusive regime, transport is extremely stable towards decoherence, even outside the critical point. Last but not least, we were able to find a {\it universal} expression for $D$, valid in the coherent diffusive regime, depending only on a single physical parameter: the ratio between the elastic scattering and the decoherence time. 

\section{The HHAA Model}
The HHAA model~\cite{AuAd80}, Fig.~\ref{fig:LET:intro}a, describes a linear chain with hopping amplitude \(J\) among sites $\ket{n}$ at distance $a$ modulated by a local potential $\varepsilon_n$, according to the Hamiltonian:
\begin{equation}
\mathcal{H}=\sum_n- J(\ket{ n}\bra{n+1}+\ket{n+1}\bra{n})+\varepsilon_n\ket{n}\bra{n},
\label{eq:HHAA-Ham}
\end{equation}
where $\varepsilon_n=W\cos(2\pi q na+\theta)$, $q=q_g={(\sqrt{5}-1)}/{2a}$ and $0<\theta<2\pi$ is a random phase over which we average in numerical simulations.
Other values of $q$ are discussed in the Appendix~\ref{sup:ChangeQ}. Contrary to the Anderson 1D model, the HHAA model presents a phase transition as the eigenstates are extended for $0\leq W<2J$ and localized for $W>2J$~\cite{AuAd80}. A notable trait is that the MIT occurs exactly at $W=2J$ in the whole spectrum and that all eigenstates have the same localization length $2\xi=a/\ln[W/{2J}]$ for $W>2J$. 

The presence of a local white-noise potential is described by the Haken-Strobl (HS) model~\cite{HaStr73}, widely used for excitonic transport. The environment is described by stochastic and uncorrelated fluctuations of the site energies $V(t)=\sum_n \varepsilon_n(t)\ket{n}\bra{n}$, with $\braket{\varepsilon_i(t)}=0$ and $\braket{\varepsilon_n(t)\varepsilon_m(t')} = \hbar \gamma_\phi \delta_{nm} \delta(t-t')$. The dynamics can be described by the Lindblad master equation:
\begin{equation}
\label{eq:complet Master ec}
\dot\rho={\cal L}[\rho] =-\frac{\mathrm i}{\hbar} \left[ \mathcal{H}, \rho\right] -\frac{\gamma_\phi}{2\hbar} \sum_{n=1}^N \left[ \ket{n} \bra{n}, \left[\ket{n} \bra{n},\rho\right]\right],
\end{equation} 
where $\gamma_{\phi}/\hbar$ is a temperature related dephasing rate. This is a good approximation when the thermal energy is on the same order of the spectral width of the system, as it happens in many biological systems~\cite{2mohseni2008environment,schulten2012}. 
It induces a diffusive spreading of the excitation in the infinite size limit of tight-binding models~\cite{moix2013coherent}. Notably, the HS master equation leads, at infinite times, to a stationary equally probable population on all sites~\cite{2spano1991cooperative}. 

Solving the master equation requires to handle \(N^2\times N^2\) matrices. To overcome this limit we use the~\textit{quantum-drift} (QD) model~\cite{FePa15}, an approach conceived as a realization of the Büttiker's local voltage probes~\cite{Pa91} in a dynamical context. Here, the system wave function follows a Trotter-Suzuki dynamics with local collapse processes represented as local energies fluctuating according to a Poisson process. This yields local energies with a Lorentzian distribution of width \(\gamma_{\phi}/2\) (for details see Appendix~\ref{SM:QD}), allowing us to handle more than $10^4$ sites.

The diffusion coefficient $D = \sigma^2(t)/(2t)$ is computed numerically through the variance $\sigma^2(t)=a^2\left[ \sum_n \rho_{n,n}(t) n^2 -(\sum_n \rho_{n,n}(t) n)^2 \right]$ starting from a local initial excitation in the middle of the chain. Our results have been confirmed using the Green-Kubo approach developed in~\cite{moix2013coherent}, see also Appendix~\ref{SM:DJianshu}.

\subsection{Coherent dynamics in HHAA.}
The initial excitation spreading is always ballistic, $\sigma_0^2(t)=v_0^2 t^2$, with a velocity $v_0^2=2a^2(J/\hbar)^2$. In absence of dephasing the long time behavior of the variance $\sigma^2_0(t)$ can be computed analytically, see Appendix~\ref{HHAA-SecondMoment}, in the three regimes: i) For $W<2J$ the spreading is still ballistic, but with a different \textit{mean group velocity}: $u^2={a^2|2J-W|^2}/{2\hbar^2}$, see solid green (upper light-gray) curve in Fig.~\ref{fig:LET:intro}b. ii) For $W>2J$, localization occurs and the variance saturates at the value $\sigma_0^2(\infty)= 2\xi^2=2a^2(2\ln(W/2J))^{-2}$, see solid blue (bottom dark-gray) curve in Fig.~\ref{fig:LET:intro}b. iii) At the MIT for $W=2J$, the variance grows diffusely~\cite{hiramoto1988dynamicsII,fleischmann1995quantum}, $\sigma_0^2(t)=2D_0t$, see middle red/gray curve in Fig.~\ref{fig:LET:intro}b. 
This is consistent with Ref.~\cite{purkayastha2018anomalous}, where deviations from a diffusive regime are shown not to affect the variance at criticality up to extremely large system sizes, where a weak super-diffusive dynamics will emerge at very large times.

The diffusion coefficient $D_0= (v_0^2 \tau_W )/2$ at the MIT depends on both the initial velocity $v_0$ and the \textit{mean elastic scattering time} \(\tau_W\) over which local inhomogeneities manifest themselves in the dynamics of a \textit{local} excitation: 
\begin{equation}
\tau_W={\hbar}/{\Delta E}, \text{ where } (\Delta E)^2 = \langle{(\varepsilon_{n}-\varepsilon_{n+1})^2}\rangle/{2}
\label{LET:eq:tw}
\end{equation}
where $\varepsilon_n=\mathcal{H}_{n,n}= \langle n|\mathcal{H}|n\rangle$ and $\langle\cdots\rangle$ represents the average over all Hamiltonian diagonal elements 
(when considering disordered models, $\langle...\rangle $ also includes average over disorder).
For the HHAA model we have
$(\Delta E)^2 = W^2(1-\cos{(2\pi qa)})/2$, and consequently, $D_0=a^2J^2/(\hbar\Delta E)$ which we checked to be in very good agreement with the numerical results at the MIT, see Fig.~\ref{fig:LET:intro}b and Appendix~\ref{sup:DtDeduction}~\&~\ref{sup:ChangeQ}.

\subsection{Decoherence in the HHAA model.}
When the system is in contact with an environment, the time-dependent fluctuations of the site energies affect the dynamics, inducing a diffusive behavior. In Fig. \ref{fig:LET:intro}b we show (symbols), for $W<2J$ and $W>2J$, how the dynamics become diffusive after a time $\tau_{\phi}\approx\hbar/\gamma_\phi$ (see vertical dotted line). In general, the diffusion coefficient depends on the decoherence strength, apart at the MIT, where, interestingly, the dynamics remains diffusive with a diffusion coefficient very close to $D_0$ as in absence of noise. 

As the decoherence strength increases, $D$ decreases in the extended regime, while in the localized regime $D$ reaches a maximum, as clearly shown in Fig.~\ref{fig:LET:intro}c. Remarkably, at the MIT, \(D\) is almost independent from decoherence up to $\gamma_{\phi}^c=2\hbar/\tau_{W}$, see red squares and vertical dotted line in Fig.~\ref{fig:LET:intro}c.
Plotted as a function of the on-site potential strength, the diffusion coefficient curves for different decoherence strengths, intersect at $W=2J$, suggesting the independence of decoherence precisely at the MIT (See Fig.~\ref{fig:Sup:DvsGa01} of Appendix~\ref{SM:DJianshu}).

In order to understand the exact dependence of $D$ on $\gamma_{\phi}$ we apply a quantum collapse model for the environmental noise. The latter can be assimilated to a sequence of measurements of the excitation's position~\cite{FePa15}, inducing local collapse that leads to a random walk~\cite{Pa91}. Then $D$ can be readily determined from $\sigma^2_0(t)$ as:
\begin{equation}
D \simeq \int_{0}^{\infty} d t_{i} p\left(t_{i}\right) \sigma_{0}^{2}\left(t_{i}\right)/(2\tau),
\label{Eq:DSigama0-gen}
\end{equation}
where $p(t_i)$ is the probability density of measurement at time $t_i$ and $\tau=\int_{0}^{\infty} d t_{i} t_i p\left(t_{i}\right)$, details are given in Appendix~\ref{SM:AnaliticalD-Derivation}. Since the HS model corresponds to a Poisson process for the measurement collapses~\cite{FePa15}, $p(t_i)=e^{-t_i/\tau_\phi}/\tau_\phi$. From $\sigma^2_0(t)$ obtained in absence of dephasing and integrating numerically Eq.~(\ref{Eq:DSigama0-gen}) we obtain results in excellent agreement with numerical data, see black curves in Fig.~\ref{fig:LET:intro}c. In Appendix~\ref{SM:subsec:DLimits} we use formalism to derive analytical expressions for $D$ in the HHAA model at the low and strong decoherence limits, these results are consistent with the expressions derived in Ref.~\cite{moix2013coherent,cao2016quantum}. 
Aditionally, from Eq.~(\ref{Eq:DSigama0-gen}), assuming a diffusive dynamics in absence of dephasing $\sigma^2_0(t)=2 D_0 t$, we can get immediately that $D=D_0$, i.e. it is independent of $\gamma_\phi$.

\begin{figure}[t]
\centering
\includegraphics[width=1\columnwidth]{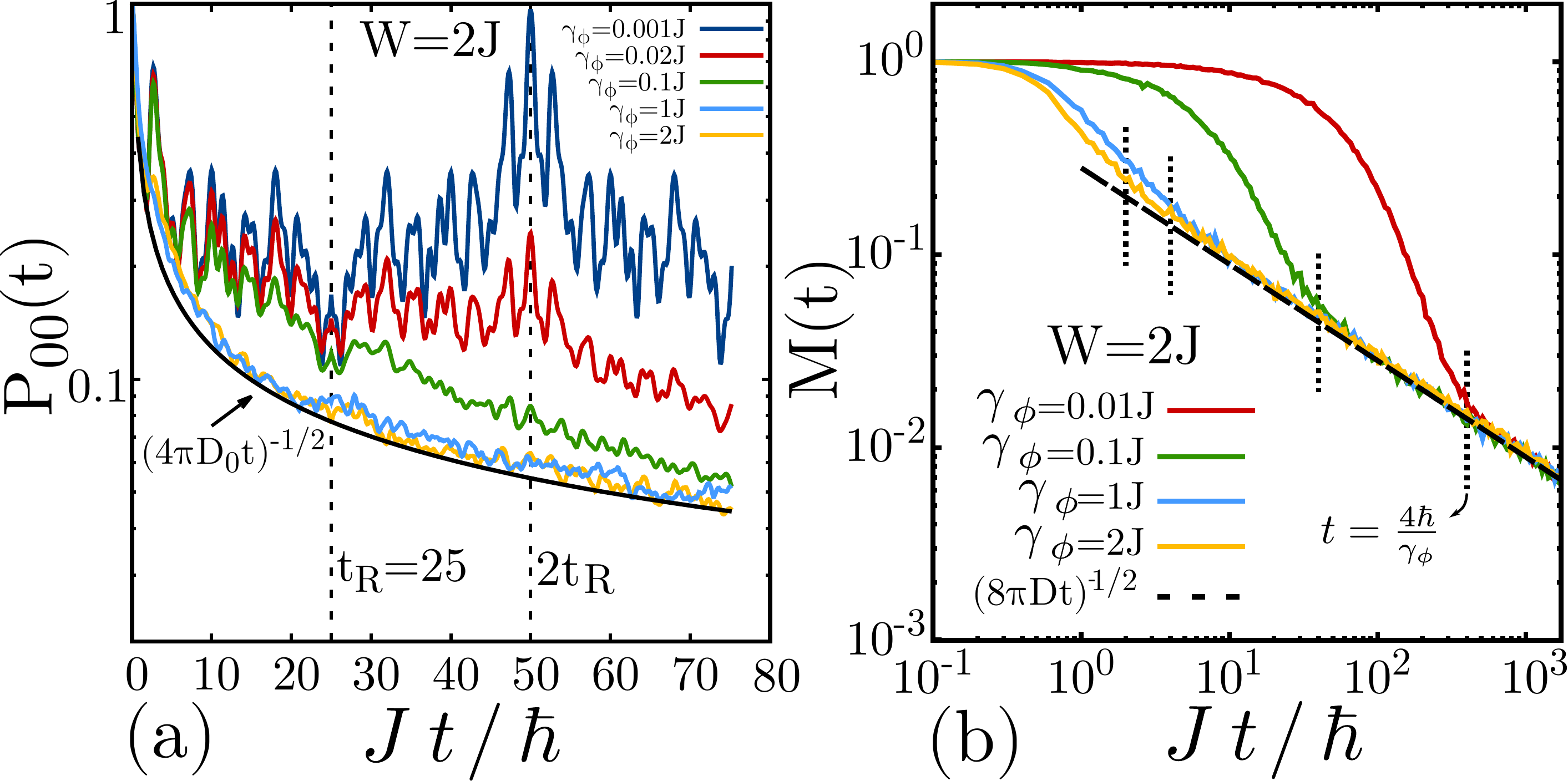}
\caption{\textbf{(a)} Probability of finding the excitation in the initial site $P_{00}(t)$ for a HHAA chain for a system evolving with $\mathcal{L}$ until time $\tau_R=25$ (first vertical dashed line) when the sign of the Hamiltonian is inverted (i.e. it continues evolving with $\mathcal{L}^\dagger$). The Loschmidt echo occurs at $P_{00}(t=2\tau_R)\equiv M(t=\tau_R)$ (second vertical dashed line) and corresponds to the purity. Different colors (gray-tones) distinguish decoherence strengths. For the stronger ones the echo is not evident and $P_{00}(t)$ approaches a diffusive dynamics (black curve).
\textbf{(b)} Loschmidt echo decay $M(t)$ for different $\gamma_\phi$ at the critical point computed with the QD. The dashed line is a prediction based on the coherent diffusion coefficient resulting from the Hamiltonian dynamics. Vertical dotted lines show $t=\frac{4\hbar}{\gamma_\phi}$. All data with $q=(\sqrt{5}-1)/2$, $J=1$, $\hbar=1$, $W=2J$ and $N=1000$.}
\label{fig:LET:02}
\end{figure}

\subsection{Loschmidt echo/purity decay in the HHAA model.}
The robustness of the wave packet spreading at the MIT, leads to the question of how the hidden decoherent processes could be unraveled from a diffusive behavior. The natural answer appears by studying how reversibility is affected by decoherence. A coherent diffusive dynamics can be reversed by changing the sign of the Hamiltonian, but the presence environment destroys the coherence that allows a perfect reversibility. This can be experimentally studied through the decay of the Loschmidt echo or purity~\cite{JaPa01,Cu-Zu03}. Purity, $M(t)=\Tr [\rho(t)^2]$, has been widely used to measure how decoherence affects a system since $M(t)\equiv 1$ for a pure state while $M(t)<1$ for a mixed state. The Loschmidt echo (LE) results from reverting the Hamiltonian part of a dynamics at a time $t_R$ through the change in the overall sign of the Hamiltonian while the environmental noise is kept active. The return probability to the initial state $P_{00}(t)$ tends to show a revival at $2t_R$. In the Appendix~\ref{SM:purity} we show that both definitions, purity and LE, are indeed equivalent, \textit{i.e.} $P_{00}(2t_R)=M(t_R)$. This allows an efficient computation using the Quantum Drift method.

Fig.~\ref{fig:LET:02}a shows the probability of finding the excitation in the initial site $P_{00}(t)$ as a function of total evolution time for different values of $\gamma_\phi$, the excitation evolves with $\mathcal{L}$, see Eq.~(\ref{eq:complet Master ec}), until time $\tau_R$ (first vertical dashed line) when the sign of the Hamiltonian is inverted (i.e. for $t>t_R$ it continues evolving with $\mathcal{L}^\dagger$). The Loschmidt echo occurs at $P_{00}(t=2\tau_R)\equiv M(t=\tau_R)$ (second vertical dashed line). However, for strong decoherence the echo is missed among the statistical fluctuations. In this case the value at $t=2\tau_R$ is mainly determined by a ``forward dynamics'' $P_{00}(t)\sim 1/\sqrt{4\pi D t}$. 

Fig.~\ref{fig:LET:02}b shows the LE/Purity $M(t)$ as a function of time. For $t\gtrsim 4\hbar/\gamma_\phi$ the initial exponential LE/Purity decay, characterized by the decoherence rate $2\gamma_\phi$, becomes a power law determined only by the diffusion coefficient: $M(t)\sim 1/\sqrt{8\pi D t}$. 
This regime is a consequence of the impossibility of reverting the spreading of the excitation beyond a time scale $2\hbar/\gamma_\phi$. Thus the LE decay only senses the spreading dynamics, which for $W=2J$, is robust against decoherence. For much stronger decoherence strengths ($\gamma_\phi \gg \gamma^c_\phi$) $D$ is in the quantum Zeno regime generating a slow decay in the purity according to $D\propto 1/\gamma_\phi$.

All the above results could be experimentally tested in Yb cold atoms in a 1D optical lattice where the HHAA model was already implemented~\cite{Nroati2008anderson,fallani2007}. Local decoherence could be imposed by time-dependent white noise fluctuations that exploit speckle patterns uncorrelated in time and space.

\begin{figure}
\centering
\includegraphics[width=1\columnwidth]{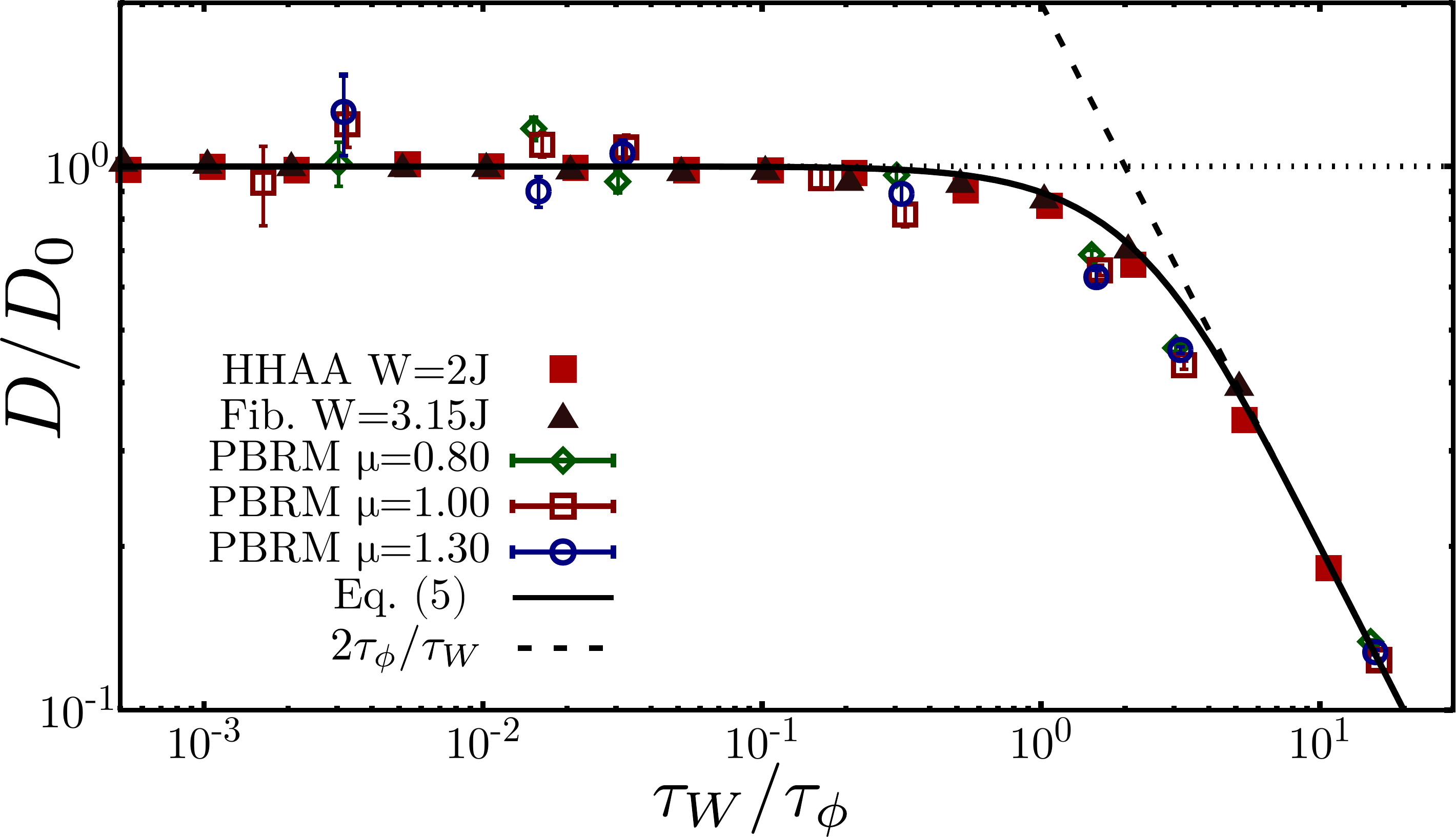}
\caption{Normalized diffusion coefficient $D/D_0$ vs. renormalized decoherence strength $\tau_W/\tau_\phi$ ($D_0=D(\gamma_\phi=0$)). Symbols obtained from QD dynamics: \textbf{i)} the HHAA chain at criticality (red squares), \textbf{ii)} the Fibonacci chain (dark-red triangles) and \textbf{iii) }the PBRM model in the extended phase (green hollow-diamonds), at the critical point (red hollow-squares) and in the localized phase (blue hollow-circles). The solid curve is the universal Eq.~\eqref{Eq:AnaliticaFromPoisson} while the dashed-black line is the limit of $\tau_W/\tau_\phi>2$. Horizontal dotted line is $D=D_0$. For the HHAA and the Fibonacci chains, $\tau_W$ and $D_0$ were computed analytically,~Eq.~\eqref{LET:eq:tw}. For the PBRM model $b=0.01$ and $D_0$ results from a fitted $\tau_W$.}
\label{fig:LET:03}
\end{figure}

\section{Stability of quantum diffusion: The Fibonacci and PBRM models.}
In order to understand whether the robustness found in the HHAA model at criticality is due to the presence of a critical point or to the presence of a diffusive dynamics, we also studied other two models: A) the Fibonacci chain~\cite{Fibonacci1983,Mer85,hiramoto1988dynamicsII} where there is no MIT but transport changes smoothly from super-diffusive to sub-diffusive as the strength of the on-site potential is varied; B) The PBRM model~\cite{mirlin_transition_1996} which presents a MIT and a diffusive second moment in a finite range of parameters around the MIT (see Appendix~\ref{SM:TestModels}). Since this model incorporates topologically different Feynman pathways it is often considered a \(1^+\)D system.

The Fibonacci model~\cite{Mer85} is described by the Hamiltonian \eqref{eq:HHAA-Ham}, with on-site energies alternating between two values as in binary alloy models: $\varepsilon_n=W(\lfloor (n+1) q_g^2\rfloor-\lfloor n q_g^2\rfloor)$, where $\lfloor...\rfloor$ is the integer part. This model has no phase transition and the variance of an initial localized excitation grows in time as $\sigma^2_0(t)\propto t^{\alpha}$ where $0<\alpha<2$ depends continuously on the on-site potential strength~\cite{varma2019diffusive,lacerda2021dephasing}. On the other side, the Hamiltonian matrix elements for the PBRM model are taken from a normal distribution with zero mean, and variance: $\braket{ |\mathcal{H}_{ij}|^2} = {1}/( {2+2(|i-j|/b)^{2 \mu}})\text{  if  } i\neq j,$ and $\braket{|\mathcal{H}_{ii}|^2} =1$ for on-site energies. The model has a critical interaction range ($\mu=1$), characterized by a multi-fractal nature~\cite{mirlin_multifractality_2000,cuevas_anomalously_2001,mendez2012multifractal}, for all values of $b$ where the system switches from extended ($\mu<1$) to localized ($\mu>1$) eigenstates~\cite{mirlin_transition_1996}. Nevertheless, small values of b allows to study the dynamics without resorting to unmanageable large systems. In this model, we have found a diffusive excitation spreading in absence of decoherence, not only at the critical point but in a much broader range of $\mu$ values $1/2 < \mu < 3/2$. Note that even for $1\le \mu \le 3/2$, the saturation value of the variance grows with the system size, thus allowing a diffusive-like spreading in the infinite size limit, see Appendix~\ref{SM:TestModels}. This sounds counter-intuitive since for $1\le \mu \le 3/2$ we are in the localized regime if the participation ratio of the eigenstates is used as a figure of merit for localization~\cite{mirlin_transition_1996}. This peculiarity is due to the long-range hopping present in this model.

\subsection{Universal stability towards decoherence.} 
As we discussed below Eq.~\eqref{Eq:DSigama0-gen}, if the coherent dynamics is diffusive at all times, then $D=D_0$ for all decoherence strengths. On the other hand, in the more realistic case, where
an initial ballistic dynamics, $\sigma^2_0(t)=v^2_0t^2$ for $t < \tau_W$, is followed by a diffusive spreading $\sigma^2_0(t)=2D_0t$, Eq.~\eqref{Eq:DSigama0-gen} yields (see details in Appendix~\ref{SM:AnaliticalD-Derivation}):
\begin{equation}
{D(x)}/{D_0}=\left[{2}/{x}-\left(1+{2}/{x}\right)e^{-x}\right],
\label{Eq:AnaliticaFromPoisson}
\end{equation}
where $x=\tau_W/\tau_{\phi}$.
This expression captures the dependence of $D$ with large and small values of $\tau_W/\tau_
{\phi}$. For $\tau_W/\tau_\phi \ll 1$, the diffusion coefficient $D\approx D_0(1-\frac{1}{6}(\frac{\tau_W}{\tau_\phi})^2)$, while for $\tau_W/\tau_{\phi} \gg 1$, we enter the strong quantum Zeno regime and $D/D_0 \approx 2\tau_{\phi} / \tau_W$. 

Eq.~\eqref{Eq:AnaliticaFromPoisson} represents our main result. As one can see, it depends only on a single parameter, i.e. the ratio between the mean elastic scattering time and the decoherence time. Thus, it describes {\it universally} any 1D quantum mechanical model characterized by a coherent diffusive dynamics, independently of the details of their microscopic dynamics.
Our analytical results have been confirmed numerically in 
Figure~\ref{fig:LET:03}, where the normalized diffusion coefficient $D/D_0$ is shown for the HHAA, Fibonacci, and PBRM models, focusing only on the diffusive-like coherent dynamics regime, where $D_0$ is well defined.
The universal behavior predicted by Eq.~\eqref{Eq:AnaliticaFromPoisson} is in excellent agreement with the numerical results for all models.

The fact that a coherent diffusive quantum dynamics is extremely robust to the environmental noise is in striking contrast with what one would expect considering scattering (with a time scale $\tau_W$) and environmental noise (with a time scale $\tau_{\phi}$) as two independent Poisson's processes. In this case, the two processes can be thought of as a single Poisson's process with a time scale $1/\tau=1/\tau_W+1/\tau_\phi$. Thus, for small values of $\tau_W/\tau_{\phi} \ll 1$, we have \(D\approx D_0(1-\tau_W/\tau_\phi)\), in contrast with the quadratic correction present in Eq.~\eqref{Eq:AnaliticaFromPoisson}. Our findings are also in contrast with standard results in classical systems where the diffusion coefficient for the dynamics in presence of external noise is the sum of the diffusion coefficients given by the two processes~\cite{chaos1984}. 

\section{Discussion}
By studying quantum transport in three paradigmatic 1D models, all of them able to support a quantum diffusive-like regime, we found a striking stability of transport towards local decoherent processes which also shows up in the purity/Loschmidt echo decay. This stability originates in the diffusive nature of the coherent quantum dynamics and it manifests itself in the fact that the diffusion coefficient is largely independent of the decoherence strength (i.e. approximately equal to the diffusion coefficient in absence of decoherence) as long as the decoherence time is longer than the mean elastic scattering time. Moreover, in the coherent diffusive regime, we analytically derived a universal law in which the diffusion coefficient depends on a single parameter: the ratio between these characteristic times. We stress that this stability does not show up when a sample is in a ballistic or in a localized regime, where the diffusion coefficient is highly sensitive to decoherence. 

We expect that our results will be valid in many realistic situations, even beyond 1D systems. In many quasi-1D systems, as occurs in the PBRM model, the elastic mean-free-path may become much larger than the localization length~\cite{4beenakker1997random} and thus, the diffusion-like regime would occur in a wide range of parameters.
Therefore, even when coherent diffusion only occurs within a limited length scale, it could be enough to ensure an efficient and stable transport under environmental noise. 

Apart from the cold atomic set-up where our findings could be verified as explained above~\cite{Nroati2008anderson,fallani2007}, 
another situation that fits the above condition is the spreading of nuclear spin excitations in quasi 1D crystals~\cite{Peng23}. There, the natural dipolar long-range interactions and the disorder can be turned on and off through appropriate radio-frequency pulses enabling a switch between ballistic and a quantum diffusive regimes. Notably, the many-body terms manifest as a decoherence timescale~\cite{levstein1991spin,lozano2021ergodicity}. Further experiments could test the stability of the spin diffusion towards decoherence. Moreover, some actual conducting polymer composites, arranged in bundles with degenerate active channels, may be in the stability regime discussed here
\cite{Re_Leo13}.

Our predictions may also inspire studies of quasi-1D biological systems where robust charge or excitonic transport are functionally relevant. Among these are energy transfer and self-repair of the helical DNA structures~\cite{Sza21,OBrBar17}. There, one might hint a crucial role of excitation propagation~\cite{klotsa2005electronic} in the puzzling mechanism through which DNA transmits allosteric signals over long distances~\cite{Ros+Hof21}. 
In photosynthetic systems it is essential an efficient energy transport from the antenna complex to the reaction center followed by a temperature independent electron transfer from a chlorophyll to a distant quinone. This elicited the long standing question of whether electron transfer occurs as a coherent process through conduction bands, or through multiple decoherent tunneling hops between localized states~\cite{LePaD90,MoSar23}. Our decoherent diffusion is an alternative mechanism that deserves further study. In the antenna complex itself, there is a convergence of energy scales (i.e. the couplings, disorder, and thermal fluctuations, are roughly of the same order), that could ensure the universally robust regime we discussed. Moreover, the analysis of the spectral statistics of several biologically relevant molecules suggest that they are typically at the border between a ballistic and a localized regime~\cite{vattay2015Bio}. Indeed, some proteins, micro-tubules and RNA~\cite{Naam12,ACS2010,Microtubules2013,Ja__Raw23}, show a surprisingly robust transport against temperature induced decoherence~\cite{Nitzan,Pe__Med23}.

In summary, we give a new light to the hypothesis, promoted for biological systems~\cite{vattay2014quantum,Kau19}, that being at the edge of chaos is favorable to charge or excitonic transport. Indeed chaos can lead to diffusion~\cite{Lau87} and, hence, to a quantum dynamics extremely robust with respect to environmental noise. In perspective, it would be interesting to analyze further signatures of intrinsic quantum diffusion in realistic biological systems in order to establish the functional relevance of our findings. We conjecture that quantum diffusion is a most relevant feature of Nature's \textit{poised realm}. 

\acknowledgements{FB and GLC thank L. Fallani and J. Parravicini for suggestion of the experimental realization of decoherence in cold atomic lattices. HMP thanks D. Chialvo for introducing S. Kauffman to him. GLC thanks E. Sadurni and A. Méndez-Bermudez for useful discussions. The work of FSLN and HMP was possible by the support of CONICET, SeCyT-UNC and FonCyT. Supercomputing time for this work was provided by CCAD (Centro de Computación de Alto Desempeño de la Universidad Nacional de Córdoba). FB and GLC acknowledge support by the Iniziativa Specifica INFN-DynSysMath. This work has been financially supported by the Catholic University of Sacred Heart. GLC acknowledges support by PNRR MUR project PE0000023-NQSTI.}

\bibliographystyle{apsrev4-2}
\bibliography{biblio}

\pagebreak
\onecolumngrid
\newpage

\appendix

\section{Current.}\label{Sup:Current}

In this section we study the steady-state current through the HHAA model in presence of pumping and draining of excitation from the opposite edges of the chain, in presence of dephasing. We also derive an approximate expression of the current as a function of the diffusion coefficient.

To generate a current, excitations are incoherently pumped and drained at the chain edges. This is modeled by including additional terms in the Lindblad master equation Eq.~\eqref{eq:complet Master ec} from the main text, which becomes
\begin{equation}
\label{SM:eq:complet Master ec}
\dot\rho={\cal L}[\rho] = -\frac{\mathrm i}{\hbar} \left[\mathcal{H},\rho\right] + {\mathcal L_\phi}[\rho] +  \mathcal{L} _{p}[\rho]+ \mathcal{L} _{d}[\rho],
\end{equation}
where $\mathcal{H}$ is the HHAA Hamiltonian \eqref{eq:HHAA-Ham} from the main text, ${\cal L}_\phi=-\frac{\gamma_\phi}{2\hbar} \sum_{n=1}^N \left[ \ket{n} \bra{n}, \left[\ket{n} \bra{n},\rho\right]\right]$ is the dephasing dissipator from the main text, while the additional terms,
\begin{equation}
\mathcal{L} _{p}[\rho] =\frac{\gamma_p}{\hbar}\left(\ket{1}\bra{0}\rho\ket{0}\bra{1}-\frac{1}{2}\ket{0}\bra{0}\rho-\frac{1}{2}\rho\ket{0}\bra{0}\right),
\end{equation}
and
\begin{equation}
\mathcal{L} _{d}[\rho] =\frac{\gamma_d}{\hbar}\left(\ket{0}\bra{N}\rho\ket{N}\bra{0}-\frac{1}{2}\ket{N}\bra{N}\rho-\frac{1}{2}\rho\ket{N}\bra{N}\right),
\end{equation}
are two operators modeling the pumping on the first site ($\ket{1}$) and draining from the last site ($\ket{N}$). Here $\ket{0}$ is the vacuum state, where no excitation is present in the system~\cite{botzung2020dark,chavez2021disorder}. For simplicity, here the pumping and draining rates are set to be equal in magnitude ($\gamma_p=\gamma_d$). From solving Eq.~(\ref{SM:eq:complet Master ec}) at the steady-state (${\cal L}[\rho_{ss}]=0$) one can compute the stationary current,
\begin{equation}
\label{eq: steady current}
I_{ss}=\frac{\gamma_d}{\hbar}\bra{N}\rho_{ss}\ket{N} \, .
\end{equation}
with $\rho_{ss}$ being the steady-state density operator~\cite{chavez2021disorder,botzung2020dark}. 

\subsection{Steady-state current: Average transfer time method.}

Since the master equation approach discussed above is numerically expensive, for large $N$ we use the \textit{average transfer time} method (ATT), as described in \cite{chavez2021disorder}. The average transfer time $\tau$ is defined as
\begin{equation}
 \tau= \frac{\gamma_d}{\hbar}\int_0 ^{\infty} t\braket{N \vert \exp\left(-\mathcal{L}_{\rm eff}t\right)\rho(0)\vert N} {\rm dt}
= \frac{\gamma_d}{\hbar} \braket{N \vert \mathcal{L}_{\rm eff}^{-2}\rho(0)\vert N}\,.
 \label{SM:eq:TauLind}
\end{equation}
where ${\cal L}_{\rm eff}$ is the one from Eq.~\eqref{SM:eq:complet Master ec} \emph{without pumping}.
In \cite{chavez2021disorder} it has been proved that the steady-state current determined from the master equation \eqref{SM:eq:complet Master ec} \emph{in absence of dephasing} depends only on the average transfer time, namely
 \begin{equation}
 \label{eq: current-tauphi}
 I_{ss}=\frac{\gamma_p}{\gamma_p \tau + \hbar}\,.
 \end{equation}
We have numerically verified that Eq.~\eqref{eq: current-tauphi} is valid also in presence of dephasing, so in the following we use it due to its lower numerical complexity together with a heuristic construction, detailed here below. 

\begin{figure}[t]
\centering
\includegraphics[height=6 cm]{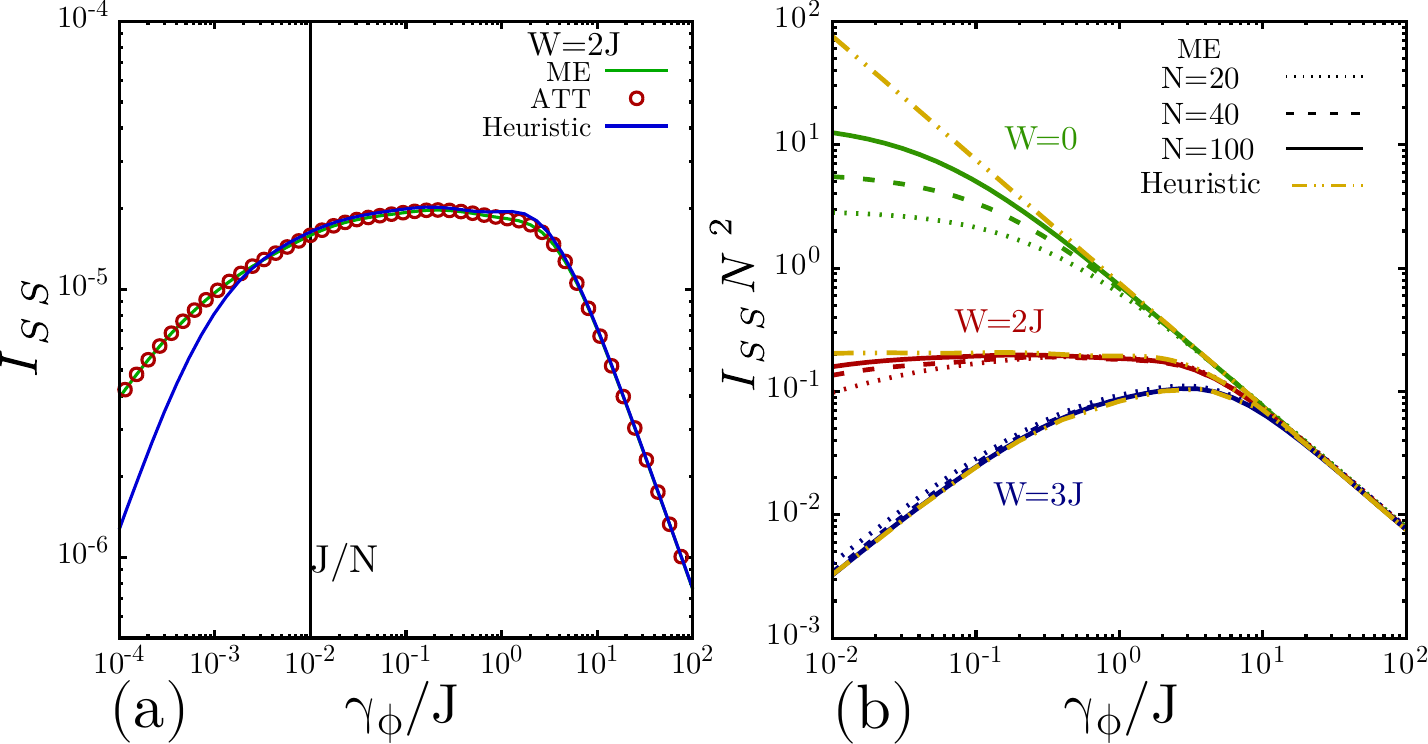}
\caption{\textbf{(a)} Steady-state current vs. $\gamma_{\phi}/j$ for the HHAA model for $N=100$ and $W=2J$ obtained with three different methods: (i) Master equation (ME, green (light-gray) curve), (ii) average transfer time method (ATT, red circles) and (iii) heuristic expression (Eq. \eqref{eq: analytic-tau}, blue (dark-gray) curve).\textbf{ (b)} Rescaled steady state current $I_{SS}N^2$ as a function of dephasing ($\gamma_\phi/J$) in the extended, critical, and localized (colors/gray-tones) regime for different system's sizes $N=\{20,40,100\}$. $I_{SS}N^2$ is calculated using the ME method and shown with different dash types depending on $N$. The diffusion coefficient based (Heuristic) estimation of the current for $N=1000$ is included with yellow (light-gray) dash-dotted curves.}
\label{SM:fig:Icomp}
\end{figure}

\subsubsection{Heuristic construction of the mean transfer time.}

The ATT method gives us the possibility to heuristically construct the mean transfer time by considering the characteristic times of dephasing-induced diffusion and draining. 

Since at equilibrium the probability of being at site $N$ is $1/N$ and the drain rate is $\gamma_d/\hbar$, we can estimate the drainage time as $\hbar N/\gamma_d$. Then, in order to determine the diffusion time, we know that an excitation moves from one site to a neighbor with an average time $a^2/(2D)$. Furthermore, the excitation moves as a random walk and the total number of steps required in 1D is $N(N-1)$. Therefore, we estimate the diffusion time as $N(N-1)a^2/(2D)$~\cite{lev2017optimal,ZhCeBoKa17}. Thus, adding the drainage time and the diffusion time we have
\begin{equation}
\label{eq: analytic-tau}
 \tau=\hbar\frac{N}{\gamma_d}+\frac{(N-1)N}{2D}a^2 \,.
\end{equation}

Figure~\ref{SM:fig:Icomp}a shows a comparison of $I_{ss}$ as a function of dephasing computed using the three different methods illustrated above here: the stationary solution of the master equation~\eqref{SM:eq:complet Master ec} (ME), the ATT method (\ref{SM:eq:TauLind}-\ref{eq: current-tauphi}), and the heuristic formula~\eqref{eq: analytic-tau}. In the latter case, the diffusion coefficient $D$ has been computed using the Green-Kubo approach [Eq.~(\ref{SM:eq:Jianshu}), Sup. Mat. \ref{SM:DJianshu}]. A general good agreement is observed between the three approaches. Deviations at small dephasing are due to the finite system size ($N=100$), for which the excitation reaches the chain edge ballistically within a time shorter than $\tau_\phi=\hbar/\gamma_\phi$.

Fig.~\ref{SM:fig:Icomp}b shows the normalized steady-state current $N^2 I_{ss}$ as a function of $\gamma_\phi$ for different $N$ in the three regimes for the HHAA model described in the main text. We observe that, as the length $N$ of the chain is increased, the behavior of the current is determined by the diffusion coefficient~Eq.~\eqref{eq: analytic-tau} (see yellow (light-gray) curves in Fig.~\ref{SM:fig:Icomp}b) where $D$ has been computed analytically for $W=0$ (Eq.~\eqref{SM:eq:DW0}) and numerically via the quantum drift approach for $W \ne 0$ and $N=1000$ (see Sec.~\ref{SM:QD}). The current decreases with dephasing in the extended regime ($W=0$) and it is enhanced in the localized regime ($W=3J$, up to an optimal dephasing), while it remains almost unaffected at the critical point ($W=2J$) up to a characteristic dephasing strength.

Although this analysis is done for the HHAA model, it should also be valid for other models with nearest-neighbor couplings, such as the Fibonacci chain analyzed in this paper. In other words, we expect that the steady-state current is mostly determined by the diffusion coefficient in such systems.

\section{The Quantum Drift Model}\label{SM:QD}

\begin{figure}
\centering
\includegraphics[height=6 cm]{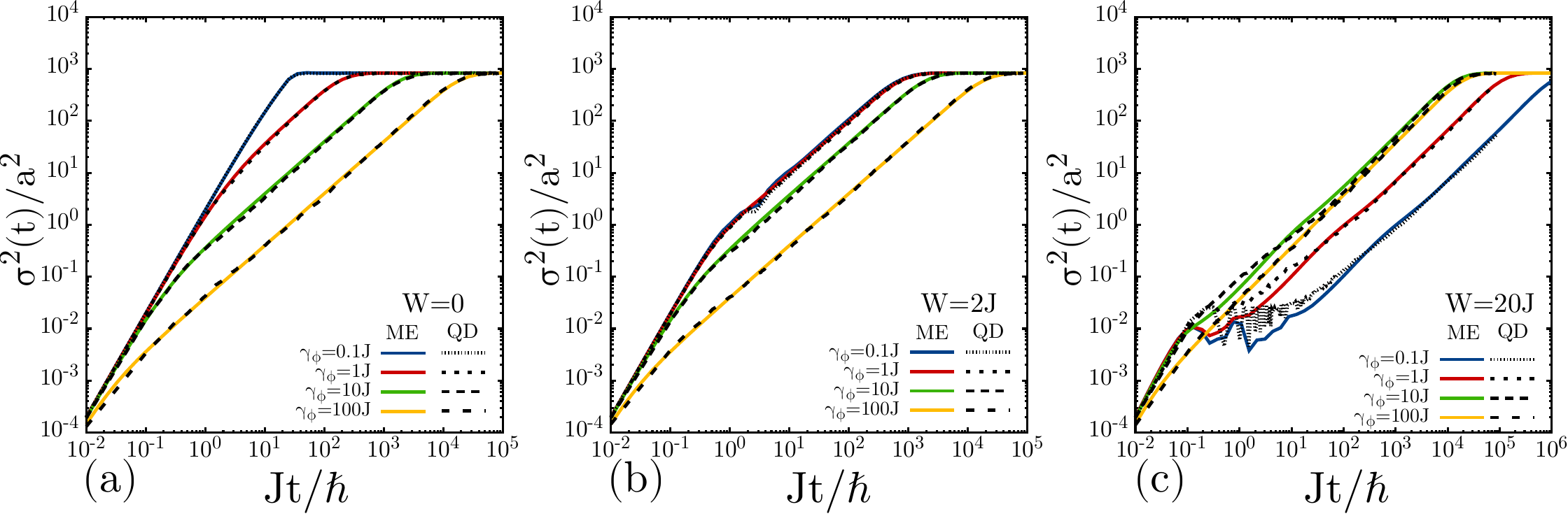}
\caption{Spreading of excitation for the HHAA model. Initial state a single site in the middle of the chain. Solid curves: Pure dephasing Haken-Strobl from a Lindblandian evolution (ME). Dashed curves: Quantum drift (QD) simulation. Colors (gray-tones) and dash-types indicate different $\gamma_\phi$ for the ME and QD data respectively. The parameters are $N=100$, \textbf{(a)} $W=0$, \textbf{(b)} $W=2J$ and \textbf{(c)} $W=20J$.}
\label{SP:Fig:Comp-QD-AA}
\end{figure}

In order to reduce the computational cost of calculation of the dynamics in presence of decoherence we use the Quantum Drift (QD) model, which only involves Trotter-Suzuki evolution of the wave-vector~\cite{FePa15,DRae83} under uncorrelated local noise. Here, the dynamics are obtained by the sequential application of unitary evolution operators to the wave-function in small time steps (d$t$). The noise/decoherence (interaction with the environment), is introduced by adding stochastic energies fluctuations on every site, $\hat{\Gamma}_\phi=\sum_n \beta_n \ket{n}\bra{n}$, uncorrelated in time (i.e. independently sampled every time step). The probability distribution of these fluctuations is a Lorentzian function,
\begin{equation}
P(\beta_n)=\frac{1}{\pi}\frac{\frac{\gamma_\phi}{2}}{\beta_n^2+(\frac{\gamma_\phi}{2})^2}.
\label{eq:prob-Lorentz}
\end{equation}
Thus, the unitary evolution in a small time step d$t$ is:
\begin{equation}
\hat{U}(\mathrm{d}t)\approx e^{{\mathrm i}\hat{\Gamma}_{\phi}\mathrm{d}t/\hbar}e^{-\mathrm{i}\hat{\mathcal{H}}\mathrm{d}t/\hbar},
\end{equation}
where $\hat{\mathcal{H}}$ is the system's Hamiltonian.
Finally, the evolved wave function at time $t=N_t\mathrm{d}t$ is:
\begin{equation}
|\hat{\psi}(t)\rangle=\prod_{j=1}^{N_t}e^{\mathrm{i}\hat{\Gamma}_{\phi}\mathrm{d}t/\hbar}e^{-\mathrm{i}\hat{\mathcal{H}} \mathrm{d}t/\hbar}|\psi(0)\rangle.
\end{equation}
The QD evolution described here is equivalent to the Haken-Strobl dephasing (Eq. \eqref{eq:complet Master ec}), see Fig.~\ref{SP:Fig:Comp-QD-AA}. As one can see there is a very good agreement between the Lindbladian and the QD evolution of the second moment of an initially local excitation for different dephasing strengths and system parameters.

\section{HHAA model: dynamics in absence of dephasing.}\label{HHAA-SecondMoment}
Here, we study the spreading of an initially localized wave packet at the center of the HHAA chain in absence of dephasing. In particular, we focus on the time evolution of the second moment $\sigma^2_0$ of the probability distribution to find the particle along the chain in absence of decoherence. As shown in the main text, in absence of decoherence and for long enough times, the second moment grows ballistically for $W<2J$, diffusively for $W=2J$ and saturates for $W>2J$\cite{fleischmann1995quantum}. 

It is known that, in the HHAA, in the localized regime the localization length of all eigenfunctions is $2\xi=a/\ln[W/{2J}]$~\cite{AuAd80,hiramoto1988dynamicsII,fleischmann1995quantum}. It follows that the wave packet probability distribution at the steady state is localized close to a site $n_0$, $P(n)=|\braket{n|\psi(t)}|^2 = \frac{1}{2\xi}(e^{-|n-n_0|/\xi})$. Therefore, the variance's saturation value will be $\lim_{t\to\infty} \sigma_0^2(t) = l^2=2\xi^2=2a^2(2\ln(W/2J))^{-2}$.
In the following we will characterize the dynamics in the different regimes.

\subsection{Extended phase.}

In the extended phase, the dynamics of the variance for very long times become ballistic, $\sigma_0^2(t)=u^2 t^2$. From the Hamiltonian [Eq.~\eqref{eq:HHAA-Ham} of the main text] in the cases $q=0$ (ordered chain) and $q=1/2$ (dimerized chain) we have proved analytically (not shown) that the velocity $u$ is directly connected with the support $B$ of the spectral bands, and we have $u^2=\frac{a^2B^2}{8\hbar^2}$. For $q=0$ there is a single band, $B=4J$ and for $q=1/2$ we have two bands, with $B=2 \sqrt{W^2+4 J^2}-2 \sqrt{W^2}$.

We here conjecture that the same expression is valid for any value of $q$ in the HHAA model. For $q$ given by the golden mean, in Ref.~\cite{thouless1983bandwidths} it was shown that $B=2|2J-W|$. Thus we have $u^2=4a^2|2J-W|^2$ and the behavior of the variance in for long times is given by:
$$\sigma_0^2(t)=\frac{a^2|2J-W|^2}{2\hbar^2}t^2.$$
These results have been confirmed numerically in Fig.~\ref{fig:LET:intro}b in the main text. 

\subsection{Critical point.}\label{sup:DtDeduction}

Here we analytically estimate the diffusion coefficient in absence of decoherence. We calculate the spreading of the wave packet $\psi(t)$ perturbatively for short times (before the scattering due to the site potential enters in the dynamics), 
so that the probability to be at site $n$ at time $t$ is: $P_n(t)=|\braket{n|\psi(t)}|^2 \simeq |\bra{n} (1-i \mathcal{H} t/\hbar)\ket{n_0}|^2$, where $n_0$ is the site where the excitation is localized initially. Defining $\mathcal{H}_{n,n_0}=\bra{n} \mathcal{H}\ket{n_0}$, and considering without lost of generality, $n_0=0$, we can write:
\begin{eqnarray}
\sigma_0^2(t)&=&a^2\sum_n P_n(t) n^2 -a^2(\sum_n P_n(t) n)^2\\&\approx&
(t/\hbar)^2a^2 \sum_n \mathcal{H}_{n,0}^2 n^2 -a^2(t/\hbar)^4\sum_{n}\mathcal{H}_{n,0}^4 n^2\\
&\approx&(t/\hbar)^2a^2 \sum_{n} \mathcal{H}_{n,0}^2 n^2=v_0^2t^2
\label{SM:eq:Sig-IniBall}
\end{eqnarray}
from which we find:
\begin{equation}
v^2_{0}=2a^2(J/\hbar)^2,
\label{eq:v0}
\end{equation}
for the HHAA since there are only nearest neighbors interactions.

We may define a time scale where the initial ballistic spreading end due to the presence of a quasi-periodic site potential of magnitude $W$. To see this effect, the perturbation expansion needs to be carry out to the 4th order: $P_n(t)=|\braket{n|\psi(t)}|^2 \simeq |\bra{n} (1-i \mathcal{H} t/\hbar-\frac{1}{2} \mathcal{H}^2 t^2/\hbar-i \frac{1}{6} \mathcal{H}^3 t^3/\hbar+\frac{1}{24}\mathcal{H}^4 t^4/\hbar)\ket{n_0}|^2$. Thus, to this level of approximation we have:
\begin{eqnarray*}
\sigma_0^2(t)/a^2 &\approx& 2J^2(t/\hbar)^2-\frac{1}{12} ((\mathcal{H}_{0,0} - \mathcal{H}_{1,1})^2+(\mathcal{H}_{0,0}- \mathcal{H}_{-1,-1})^2)J^2(t/\hbar)^4,\\
\sigma_0^2(t)/a^2 &\approx& 2J^2(t/\hbar)^2-\frac{2}{12} \langle (\mathcal{H}_{n,n} -\mathcal{H}_{n+1,n+1})^2 \rangle J^2(t/\hbar)^4. 
\end{eqnarray*}
where the energy differences squared were replaced by the average value: 
\begin{equation}
(\Delta E)^2 =\langle (\mathcal{H}_{n,n} -\mathcal{H}_{n+1,n+1})^2 \rangle =\frac{1}{N-1} \sum_{n=1}^{N-1} \frac{(\mathcal{H}_{n,n}-\mathcal{H}_{n+1,n+1})^2}{2}.
\label{eq:Delta}
\end{equation}
This definition takes into account the ``correlation'' between neighbors. For the HHAA model the average can be taken over the sites $n$ or the realizations of the potential (phase $\theta$ in Eq. \ref{eq:HHAA-Ham}). For independent random disorder (Anderson disorder), yields directly the variance of the disorder ($(\Delta E)^2 =\frac{1}{N-1} \sum_{n=1}^{N-1} \mathcal{H}_{n,n}^2$), which is the standard magnitude to calculate the disorder time scale.

The first effect of this quartic correction is to change the concavity of the $\sigma_0^2(t)$. This will happen when the second derivative of $\sigma_0^2(t)/a^2$ vanish at a time $\tau_W$, so that:
\begin{equation}
 \tau_W=\sqrt{\left(\frac{\langle (\mathcal{H}_{n,n} - \mathcal{H}_{n+1,n+1})^2\rangle}{2\hbar^2} \right)^{-1}}=\frac{\hbar}{\Delta E}.
\label{SM:eq:Gen-tauW}
\end{equation}
By replacing with the HHAA site energies, using trigonometric identities, and summing over the sites, it can be shown that $\Delta E= W \sqrt{(1-\cos{(2\pi q)})/2}$, and we have:
\begin{eqnarray}
\tau_{W}&=&\frac{\sqrt{2}\hbar}{W\sqrt{(1-\cos{(2\pi q)})}},
\label{eq:sup:tauW}
\end{eqnarray}
then, the diffusion coefficient in absence of dephasing, $D_0$, can be computed as follows:
\begin{equation}
\label{SM:eq:Dt}
D_0=\frac{v_0^2\tau_W}{2}=\frac{a^2J^2}{\hbar}\frac{\sqrt{2}}{W\sqrt{(1-\cos{(2\pi q)})}}.
\end{equation}
It is interesting to note how the correlations of the model (given by the modulation wave vector $q$) influence the scattering times and therefore the diffusion $\sigma_0^2(t) = 2 D_0 t = v_0^2\tau_W t$. 

Notice that here, the potential strength enters with a different power law than in the mean-free-time between collisions that results from the application of the Fermi Golden Rule to a Bloch state of energy \(\varepsilon\) for the \textit{uncorrelated disorder} of Anderson's model~\cite{ThKi81} \(1/\tau_{FGR}=(2\pi/\hbar)(W^2/12)N_1(\varepsilon)\) with \(N_1(\varepsilon)\propto 1/4\pi J\sqrt{1-(\varepsilon/2J)^2}\) being the density of directly connected states.

\section{Diffusion coefficient in presence of decoherence.}

\subsection{Green-Kubo formula.}\label{SM:DJianshu}

The diffusion coefficient $D$ in presence of decoherence for the Haken-Strobl model can be computed from the Green-Kubo expression, using only the eigenenergies and eigenstates of the Hamiltonian,
\begin{equation}
{\cal H} \phi^\mu = \varepsilon_\mu \phi^\mu
\end{equation}
as it has been derived in Ref.~\cite{chuang2016quantum}:
\begin{equation}
\label{SM:eq:Jianshu}
D(\Vec{u})=\frac{\hbar}{N}\sum_{\mu,\nu=1}^N \frac{\gamma_\phi}{\gamma_\phi^2+\omega_{\mu,\nu}^2}|\hat{j}_{\mu,\nu}(\Vec{u})|^2 \,,
\end{equation}
where $\gamma_\phi$ is the dephasing strength, $\omega_{\mu,\nu}=\varepsilon_{\mu}-\varepsilon_{\nu}$ is the energy difference between eigenstates $\mu$ and $\nu$, and $\hat{j}_{\mu,\nu}$ is the flux operator in the eigenbasis:
\begin{equation}
\hat{j}_{\nu,\mu}(\Vec{u})=\frac{i}{\hbar} \sum_{n,m} (\Vec{u}\cdot\Vec{r}_{n,m})\phi^{\mu *}_n \phi^{\nu}_m {\cal H}_{n,m} \,. 
\end{equation}
In the expression above, $\Vec{u}$ is a unit vector indicating the transport direction, $\Vec{r}_{n,m}$ is the vector connecting the positions of site $n$ and $m$, $\phi^\nu_n$ is the amplitude of the $\nu$ eigenstate at site $n$ and $\mathcal{H}_{n,m}=\langle n| \mathcal{H}| m \rangle$ is the coupling between $n$ and $m$ sites. In our 1D system with nearest neighbor interactions, $\Vec{u}\cdot\Vec{r}_{n,m}=m-n=\pm a$ and $\mathcal{H}_{n,m}=J(\delta_{m,n+1}+\delta_{m,n-1})$. Therefore,
\begin{equation}
\hat{j}_{\nu,\mu}=i\frac{Ja}{\hbar}\sum_{n} \phi^{\mu *}_n(\phi^{\nu}_{n+1}-\phi^{\nu}_{n-1}).
\end{equation}

\begin{figure}
\centering
\includegraphics[width=1.0\columnwidth]{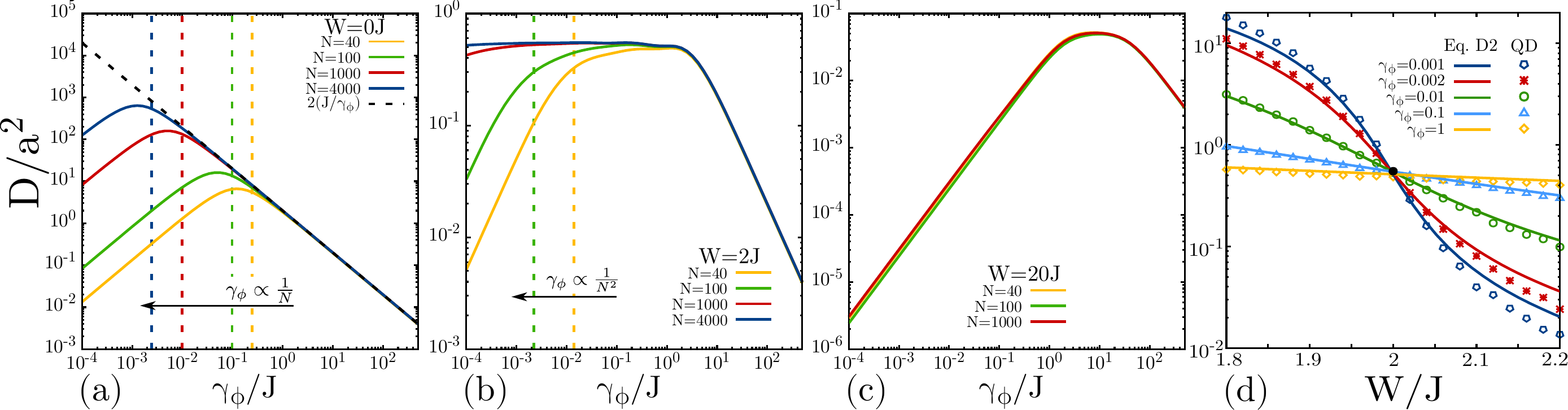}
\caption{Diffusion coefficient $D/a^2$, calculated using the Green-Kubo approach (Eq.~\eqref{SM:eq:Jianshu}), vs. $\gamma_{\phi}/J$ for the HHAA model with Haken-Stobl dephasing for different values of $N=\{40, 100, 1000, 4000\}$ (colored (gray-toned) curves). Figure \textbf{(a)} is for $W/J=0$ (metallic regime), figure \textbf{(b)} is for $W/J=2.0$ (MIT regime), and figure \textbf{(c)} is for $W/J=20$ (insulator regime). Vertical dashed lines indicate the values of $\gamma_\phi$ below which finite size effects are relevant. The dependence of this value with $N$ is shown on top of the black arrow. (d) Diffusion coefficient $D/a^2$ vs. $W/J$ for different $\gamma_\phi$ (colors/gray-tones). Curves are calculated from Eq.~\eqref{SM:eq:Jianshu}. Symbols are obtained from the Quantum Drift (QD, Appendix~\ref{SM:QD}). $N=1000$.}
\label{fig:Sup:DvsGa01}
\end{figure}

Equation \eqref{SM:eq:Jianshu} have been compared with numerical simulations using the QD approach in Figures \ref{fig:Sup:DvsGa01}d, \ref{SM:D-as-RW}, \ref{SM:DLimits}, and \ref{fig:Sup:ChQ1}. It also have been used to study the dependence with $N$ of the diffusion coefficient in various models. Figure \ref{fig:Sup:DvsGa01} shows the diffusion coefficient $D$ of the HHAA model in the three regimes as a function of the dephasing strength for different chain lengths $N$. We observe for small dephasing a clear dependence of $D$ on the system size. This is due to the fact that when dephasing is small, the excitation reaches the boundaries before diffusion can sets in. Defining the typical time scale for dephasing to affect the dynamics as $\tau_\phi=\frac{\hbar}{\gamma_\phi}$, we can estimate the dephasing strength below which finite size effects are relevant, by comparing $\tau_{\phi}$ with the time needed to reach ballistically the boundaries for the clean case ($W=0$). In the ballistic regime ($W<2J$) the value of decoherence strength below finite size effect starts to be relevant will decrease proportional to $1/N$, while in the diffusive regime ($W=2J$) with $1/N^2$ (see vertical dashed lines in Figures~\ref{fig:Sup:DvsGa01}ab). In the localized regime finite size effects are negligible if the system size is larger than the localization length. Fig.~\ref{fig:Sup:DvsGa01}d shows $D$ vs. the on-site potential strength for different decoherence strengths, curves are calculated using the Green-Kubo approach while symbols are obtained from the Quantum Drift dynamics (Appendix~\ref{SM:QD}). As one can see all curves intersect at $W=2J$, suggesting the independence of decoherence precisely at the MIT. 

\subsection{Analytical expression of the Diffusion coefficient from the coherent dynamics.} \label{SM:AnaliticalD-Derivation}

The presence of the Haken-Strobl dephasing can be thought as the system being measured by the environment\cite{AlPas07,FePa15}. This measurements occur at random times, where the times between subsequent measurements are distributed as $p(t)=e^{-t/\tau_\phi}/\tau_\phi$, with $\tau_\phi=\hbar/\gamma_\phi$. In this section, we employ this interpretation of the Haken-Strobl dephasing to obtain analytical expressions for the diffusion coefficient.

When the measurement occurs, the system has a probability distribution of being at the position $r$, $P_0(r,t,r_0,t_0)$, determined by the coherent Hamiltonian dynamics. The initial position, $r_0$ at $t_0$ will only define the center of the probability density, since the system is isotropic. This assumption is valid in the three models treated in this work unless the excitation is close to the boundaries. Consequently, $P_{0}\left(r, t, r_{0}, t_{0}\right)=P_{0}\left(r-r_0, t-t_{0}, 0,0\right)$. For simplicity we will consider $r_{0}=0$, $t_0=0$.

The probability density of measuring the system at site $r$ at time $t$ once the measurement process is included ($\widetilde{P}(r, t, 0,0)$) will be determined by the integral equation:
\begin{equation}
\widetilde{P}(r, t, 0,0)=\underbrace{P_{0}(r, t, 0,0)\left(1-\int_{0}^{t} p(t_{i}) d t_{i}\right)}_{\text{No Measurement.}}+\underbrace{\int d r_{i} \int^t_{0} d t_{i} p\left(t_{i}\right) \tilde{P}\left(r, t, r_{i}, t_{i}\right) P_{0}\left(r_{i}, t_{i}, 0,0\right)}_{\text{Measurement at } (t_{i},r_i)},
\end{equation}
which recurrently considered the probability of not being measured and of being measured several times.

To directly analyze the second moment of the distribution we multiply by $r^2$ and integrate over $r$ on both sides:
\begin{equation}
\sigma^{2}(t)=\sigma_{0}^{2}(t)\left(1-\int_{0}^{t} p(t_{i}) d t_{i}\right)+\int d r_{i} \int^t_{0} d t_{i} p\left(t_{i}\right) \underbrace{\int d r \tilde{P}\left(r,t, r_{i}, t_{i}\right) r^{2}}_{r_{i}^{2}+\sigma^{2}\left(t-t_{i}\right)} P_0\left(r_{i}, t_{i}, 0,0\right),
\end{equation}
\begin{equation}
\sigma^{2}(t)=\sigma_{0}^{2}(t)\left(1-\int_{0}^{t} p(t_{i}) d t_{i}\right)+\int_{0}^{t} d t_{i} p\left(t_{i}\right) \sigma_{0}^{2}\left(t_{i}\right)+\int_{0}^{t} d t_{i} p\left(t_{i}\right) \sigma^{2}\left(t-t_{i}\right),
\label{SM:Sigma-Volverra}
\end{equation}
where we have used the independence of the probabilities from the initial site and time. 

It can be shown by Laplace transform in Eq. \eqref{SM:Sigma-Volverra} (Appendix~\ref{SM:AnaliticalSpreading}), that for well-behaved $p(t)$ and $\sigma_0^2(t)$ (trivially fulfilled in the systems we consider), the dynamics of the variance $\sigma^2(t)$ becomes diffusive at long enough times. Therefore, in the long time limit ($t\rightarrow\infty$) we have:
\begin{eqnarray*}
\sigma^{2}(t) &\simeq& 2 D t,\\
\left(1-\int_{0}^{t} p(t_{i}) d t_{i}\right) &\simeq& 0,\\
\int_{0}^{t} d t_{i} p\left(t_{i}\right) t_i & \simeq& \tau_{\phi},
\end{eqnarray*}
and,
\begin{equation}
D = \frac{\int_{0}^{\infty} d t_{i} p\left(t_{i}\right) \sigma_{0}^{2}\left(t_{i}\right)}{2\tau_{\phi}}.
\label{SM:eq:DSigma0-gen}
\end{equation}

Then if $\sigma_{0}^{2}(t)=2 D_{0} t$ $\forall t$ the measurement process does not affect the diffusion coefficient:
\begin{equation}
D=2 D_{0} \frac{\int_{0}^{\infty} p(t_{i}) t_{i} d t_{i}}{2 \tau_{\phi}}=D_{0}. 
\end{equation}

Another physically relevant case is when the dynamics is initially ballistic up to some time $\tau_W$ followed by a diffusive dynamic.
\begin{equation}
\sigma_{0}^{2}(t)=\left\{\begin{array}{ll}
v_{0}^{2} t^{2} & \text { if } t<\tau_{W} \\
2 D_{0} t & \text { if } t>\tau_{W}
\end{array} \text { with } D_{0}=\frac{v_{0}^{2} \tau_{W}}{2}\right. \\
\end{equation}

\begin{equation}
D=\frac{1}{2 \tau_{\phi}} \left(\int_{0}^{\tau_{W}} \frac{2 D_{0}}{\tau_{W}} t^{2} p(t) d t+\int_{\tau_{W}}^{\infty} 2 D_{0} t p(t) d t\right)
\end{equation} 

Considering a Poisson process for the measurements: $p(t)=\frac{e^{-t/\tau_\phi}}{\tau_\phi}$, we have:
\begin{equation}
D(\tau_\phi)=D_0\left(\frac{2\tau_\phi}{\tau_W}-\left(1+\frac{2\tau_\phi}{\tau_W}\right)e^{-\tau_W/\tau_\phi}\right),
\label{SM:Eq:AnaliticaFromPoisson}
\end{equation}
this expression captures the dependence of $D$ for large and small values of $\tau$, so that $D\approx D_0(1-\frac{1}{6}(\frac{\tau_W}{\tau_\phi})^2)$ and $D\approx v^2_0\tau_\phi$ respectively. Note that considering a process $p_\delta(t)=\delta(t-2\tau_\phi)$, would yield $\widetilde{D}=v^2_0\tau_W=D_0$ for $\tau_\phi>\tau_W/2$ and $\widetilde{D}=v^2_0 2\tau_\phi=D_0 2\tau_\phi/\tau_W$ for $\tau_\phi<\tau_W/2$.

\subsubsection{Analytical solution for the spreading.}\label{SM:AnaliticalSpreading}

In this section we show that Eq. \eqref{SM:Sigma-Volverra} for $p(t)=e^{-t/\tau_\phi}/\tau_\phi$ generates a diffusive dynamics at long times and find analytical solutions in some paradigmatic cases. Eq. \eqref{SM:Sigma-Volverra} can be rearranged in the following form:
\begin{equation}
\sigma^{2}(t)=f(t)+\int_{0}^{t} d t_{i} p\left(t_{i}\right) \sigma^{2}\left(t-t_{i}\right),
\end{equation}
by noting that $ \left(1-\int_{0}^{t} p(t) d t\right)=e^{-t/\tau_\phi}=\tau_\phi p(t)$ and defining $f(t)=\tau_\phi g(t)+\int_{0}^{t} d t_{i}g(t_i)$ with $g(t)=\sigma_{0}^{2}(t)p(t)$.

The usual strategy to solve this type of equation is to use the Laplace's transform on the equation, $$\sigma^2_{LT}(s)=\mathcal{F}(s)+\sigma^2_{LT}(s)\mathcal{P}(s),$$ where $\sigma^2_{LT}(s)$, $\mathcal{F}(s)$ and $\mathcal{P}(s)=\frac{1}{s\tau_\phi+1}$ are the Laplace's transform of $\sigma^2(t)$, $f(t)$ and $p(t)$ respectively.

Identifying $\mathcal{G}(s)$ as the Laplace's transform of $g(t)$, we have
$\mathcal{F}(s)=\mathcal{G}(s)(\frac{s\tau_\phi+1}{s})$, and:

\begin{equation}
\sigma^2_{LT}(s)=\frac{\mathcal{F}(s)}{1-\mathcal{P}(s)}=\mathcal{G}(s)\tau_\phi \frac{(s\tau_\phi+1)^2}{(s\tau_\phi)^2}=\mathcal{G}(s)\tau_\phi\left[\frac{1}{(s\tau_\phi)^2}+\frac{2}{s\tau_\phi}+1  \right].
\end{equation}
Since the Laplace transform of $t^n u(t)$, where $u(t)$ is the step function, is $\frac{n!}{s^{n+1}}$ we observe that $\sigma^2(t)$ will be diffusive in the long time limit if $\mathcal{G}(0)$ is finite and non zero, a condition trivially fulfilled in the systems under consideration. In this case, $D=\frac{\mathcal{G}(0)}{2\tau_\phi}=\frac{\int_0^{\infty}\sigma_0^2(t)p(t)dt}{2\tau_\phi}$, as we found in Eq. \eqref{SM:eq:DSigma0-gen}.

The inverse transform of $\sigma_{LT}^2(s)$ can be carried out in several cases (for example $\sigma_0^2(t)=A_\alpha t^\alpha$), however, here we only discuss two paradigmatic cases $\sigma_0^2(t)=2D_0t$ and $\sigma_0^2(t)=v_0^2t^2$. In the first case, the diffusive spreading, we find $\sigma^2(t)=2D_0 t$, i.e. the dynamic of $\sigma_0^2$ is not affected. 

In the second case, the ballistic spreading, the solution is 
\begin{equation}
\sigma^2(t)=2 \tau_\phi  v_0^2 \left(\tau_\phi 
\left(e^{-\frac{t}{\tau_\phi }}-1\right)+t\right),
\end{equation}
which for $t\ll\tau_\phi$, $\sigma^2(t)\approx v_0^2 t^2$, maintains its ballistic behavior but becomes diffusive for $t\gg\tau_\phi$, $\sigma^2(t)\approx 2 v_0^2\tau_\phi t=2Dt$. The same expression is found when the spreading in an ordered tight-binding chain with Haken-Strobl decoherence is addressed with the Lindblad formalism \cite{Rei82,moix2013coherent}.

It is important to note that if one considers two Poisson processes, $p_1(t)=e^{-t/\tau_1}/\tau_1$ and $p_2(t)=e^{-t/\tau_2}/\tau_2$, the combined effect will be equivalent to consider only one process with $p(t)=e^{-t/\tau}/\tau$ with $\tau=\frac{\tau_1 \tau_2}{\tau_1 +\tau_2}$, the sum of inverse times scales. This is the standard result in classical systems where one considers a particle that moves with velocity $v_0$ to the left or right with the same probability after a scattering with either of the two processes. The diffusion coefficient, in this case, is, $D= v_0^2\tau=v^2_0\frac{\tau_1 \tau_2}{\tau_1 +\tau_2}=D_1 \frac{1}{1+\tau_1/\tau_2}$, which for $\tau_2\gg\tau_1$ generates a linear correction to the diffusion coefficient associated with the process $p_1$.

\subsection{Analytical expression of the Diffusion coefficient in the limit of strong and weak decoherence.} \label{SM:subsec:DLimits}

\begin{figure}
\centering
\includegraphics[height=6 cm]{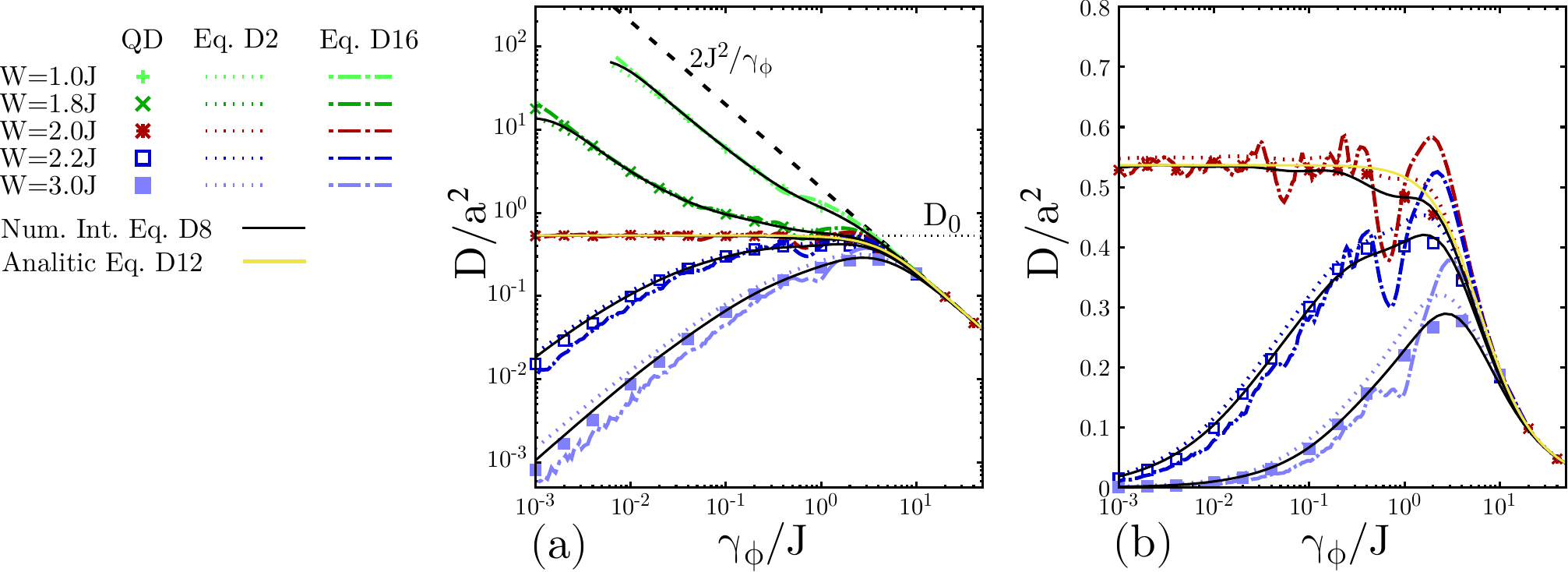}
\caption{ \textbf{(a,b)} Diffusion coefficient vs. the decoherence strength for the HHAA model. Symbols have been obtained from the time evolution, dotted curves from Eq. \eqref{SM:eq:Jianshu} [Green-Kubo expresion], dash-dotted colored (gray-toned) curves from Eq. \eqref{SM:eq:Dsigma0} [Delta process] and solid black curves from Eq. \eqref{SM:eq:DSigma0-gen} [Poisson process]. Also shown: the diffusion coefficient in the strong dephasing regime (dashed-black line) and the diffusion coefficient in absence of dephasing at the MIT ($D_0$) as a horizontal dotted line, see Eq.~(\ref{SM:eq:Dt}). The yellow curve corresponds to Eq. \eqref{SM:Eq:AnaliticaFromPoisson}. Panel (b) is the same as (a) but in linear-log. scale and excluding the extended regime data.}
\label{SM:D-as-RW}
\end{figure}

In this section, we use Eq.~\eqref{SM:eq:DSigma0-gen} and the specific dynamics of $\sigma^2_0(t)$ in the HHAA model (Appendix~\ref{HHAA-SecondMoment}), to obtain the behavior of $D$ in the limit of strong and weak dephasing.

We define the mean free path $l$ from the expectation value of the coherent spreading $l^2=\int_0^\infty \sigma_0^2(t) p(t) dt$. We compare it with a random walk analysis of the diffusion coefficient~\cite{lloyd2011quantum} which corresponds to a delta process where the system is measured by the environment at equal times $\delta t=2\hbar/\gamma_{\phi}$. The diffusion coefficient is directly determined by the coherent spreading at the dephasing time:
\begin{equation}
D=\frac{l^2}{2\delta t}= \frac{\sigma^2_0(t=\frac{2\hbar}{\gamma_\phi})}{2\frac{2\hbar}{\gamma_\phi}},
\label{SM:eq:Dsigma0}
\end{equation}
this expression, however inaccurate, can be considered a first approximation to the diffusion coefficient.

Figures~\ref{SM:D-as-RW} show the diffusion coefficient obtained from the time evolution (symbols), Eq.~\eqref{SM:eq:Jianshu} (dotted curves), Eq.~\eqref{SM:eq:Dsigma0} (dash-dotted colored-curves) and from numerical integration of Eq.~\eqref{SM:eq:DSigma0-gen} (solid black-curves). The yellow curve corresponds to Eq.~\eqref{SM:Eq:AnaliticaFromPoisson}. We observe that using a Poisson process (Eq.~\eqref{SM:eq:DSigma0-gen}) we obtain smoother results than with a Delta process (Eq.~\eqref{SM:eq:Dsigma0}) (the fluctuations produced by particular interferences are washed-out) and can be obtained at almost the same computational cost.

\begin{figure}
\centering
\includegraphics[height=6 cm]{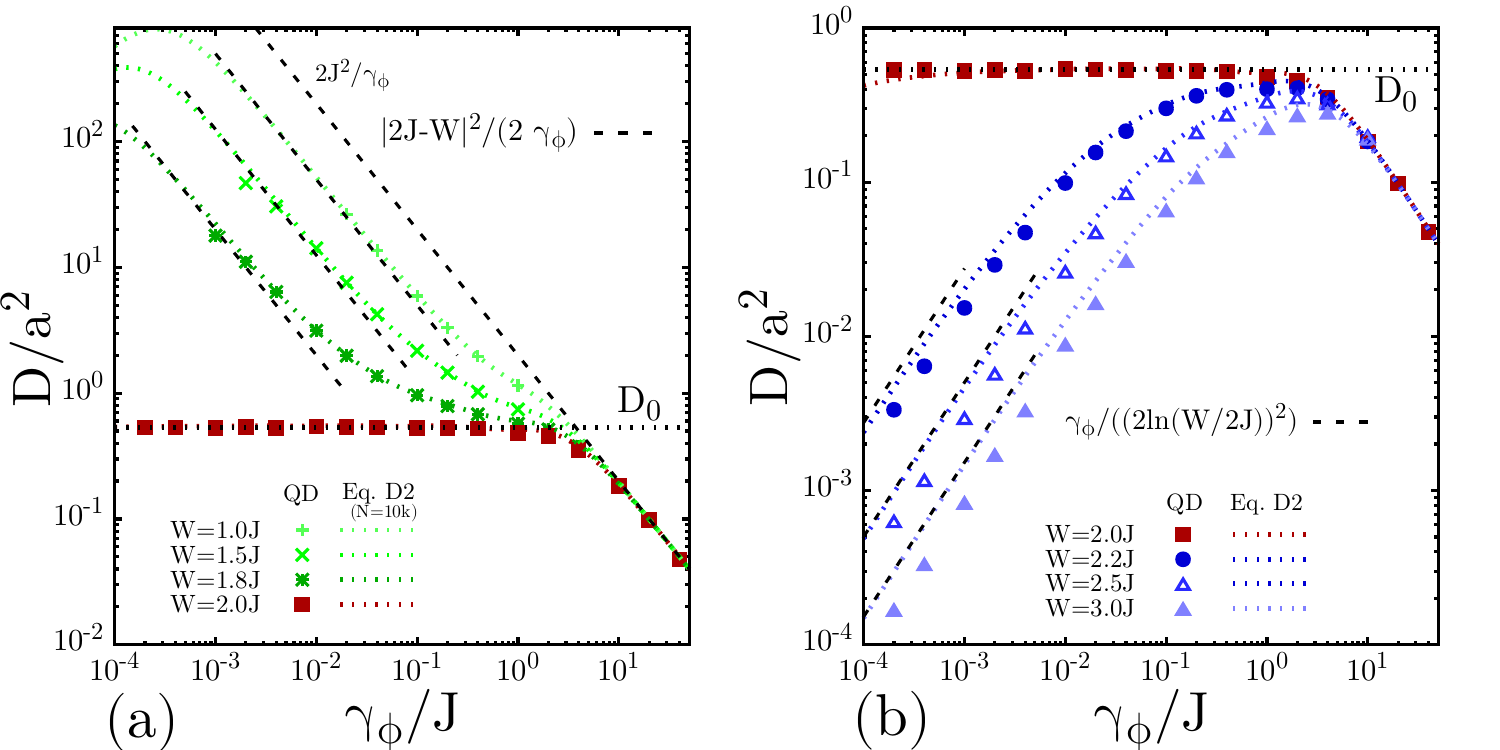}
\caption{Diffusion coefficient vs. decoherence strength for the HHAA model. Symbols have been obtained from the QD time evolution, dotted curves from the Green-Kubo formula and dashed lines represents the analytical estimations. \textbf{(a)} Extended phase, the analytical results corresponds to Eq. \eqref{SM:eq:Dbal} (dashed-black lines). \textbf{(b)} Localized phase, the analytical results corresponds to Eq. \eqref{SM:eq:Dloc} (dashed-black lines). The parameters are $N=1000$, $Q=(\sqrt{5}-1)/2$, $J=1$ and $\hbar=1$.\\}
\label{SM:DLimits}
\end{figure}

\subsubsection{Strong decoherence.}

For sufficiently large dephasing, $\gamma_\phi \gg \hbar/\tau_{W}$, the noise interrupts the dynamics before the systems notice if it is in an extended, critical, or localize phase. This is known as the strong Zeno regime. In this case, the measurement happens during the initial ballistic dynamics, where the variance grows as $\sigma_0^2(t)=2a^2\frac{J^2}{\hbar^2}t^2$. Therefore, the dynamic corresponds to a random walk with a mean free path $l^2=2a^2\frac{J^2}{\hbar^2}\delta t^2$ and a mean free time $\delta t=\frac{2\hbar}{\gamma_\phi}$. Thus, the diffusion coefficient is:
\begin{equation}
D=\frac{1}{2}\frac{2a^2J^2}{(\gamma_\phi/2)^2}\frac{\gamma_\phi}{2\hbar}=\frac{2a^2J^2}{\hbar\gamma_\phi}.
\label{SM:eq:DW0}
\end{equation}

The same result is obtained with the Poisson process $p(t)=e^{-t/\tau_\phi}/\tau_\phi$. This result is valid for all $\gamma_\phi$ for an infinite clean chain ($W=0$)~\cite{Rei82}, since in that case $\tau_W \to\infty$. Notice that Eq. \ref{SM:eq:DW0} is also valid in the presence of correlated noise (e.g. Binary and Gaussian processes), which has been shown to involve only a renormalization of the decoherence strength for short correlation times\cite{FePa15,amir2009classical}.

\subsubsection{Extended phase \texorpdfstring{($W<2J)$}{TEXT}.}

 For sufficiently small dephasing strength (depending on how close we are to the MIT), the system enters the long-time ballistic regime where $\sigma^2_0(t)=\frac{a^2|2J-W|^2}{2\hbar^2}t^2$ from which we have: 
 \begin{equation}
D=\frac{a^2|2J-W|^2}{2\hbar\gamma_\phi}.
\label{SM:eq:Dbal}
\end{equation}
Note that as we approach the MIT our estimate is valid for a smaller and smaller dephasing strength since the system enters the ballistic regime at larger times. Using the Poisson process $p(t)$ and Eq.~\eqref{SM:eq:DSigma0-gen} we obtain the same results. In Fig.~\ref{SM:DLimits}a we compare the diffusion coefficient obtained from the numerical simulations (symbols) with the analytical approximation Eq.~\eqref{SM:eq:Dbal}.

\subsubsection{MIT \texorpdfstring{$(W=2J)$}{TEXT}.}

At the critical point, for $t>\tau_W$ the dynamic is diffusive and the variance is linearly dependent on the measurement time $\sigma^2_0(\delta t)=2 D_0 \delta t$. Given that we have $l^2=2D_0 \delta t$, provided that $\gamma_\phi<2\hbar/\tau_W$, and $D= l^2/(2\delta t)$ we obtain: 
\begin{equation}
D=\frac{2D_0\delta t}{2\delta t}=D_0,
\end{equation}
i.e. a diffusion coefficient independent of the dephasing.

This was shown to be exact for an always diffusive dynamic in Appendix~\ref{SM:AnaliticalD-Derivation}. On the other hand when we consider a ballistic dynamics for short times and a Poisson measurement process some corrections appear.

\subsubsection{Localized phase \texorpdfstring{$(W>2J)$}{TEXT}.}

For sufficiently small dephasing strength (depending on how close we are to the MIT), the system gets localized with a localization length $\xi=l/\sqrt{2}$ before dephasing sets in. So, considering $\sigma^2_0=l^2$ in Eq. \ref{SM:eq:DSigma0-gen}:
\begin{equation}
\sigma^2(t)=\frac{l^2}{\tau_\phi}t=l^2\frac{\gamma_\phi}{\hbar}t=2\xi^2\frac{\gamma_\phi}{\hbar}t.
\label{SM:eq:SigLoc}
\end{equation}
This limit is also found in Ref. \cite{chuang2016quantum} from Eq. \ref{SM:eq:Jianshu}. Since in the HHAA model $2\xi^2=2a^2(2\ln(W/2J))^{-2}$, the diffusion coefficient is:
\begin{equation}
D=\frac{\xi^2\gamma_\phi}{\hbar}=\frac{a^2\gamma_\phi}{(2\ln(W/2J))^{2}\hbar}.
\label{SM:eq:Dloc}
\end{equation}
The analytical result is shown in Fig.~\ref{SM:DLimits}b compared with the numerical results. We observe a small discrepancy with the above formula, rooted in the fact that the numerically found $l^2$ is slightly smaller than the theoretical one. 

Notice that, in contrast with the other regimes, the delta and Poisson process do not yield the same expression (the use of a delta process would underestimates the diffusion coefficient by a factor of two).

\section{HHAA model with different values of \texorpdfstring{$q$}{TEXT}.}\label{sup:ChangeQ}

\begin{figure}
\centering
\includegraphics[height=12 cm]{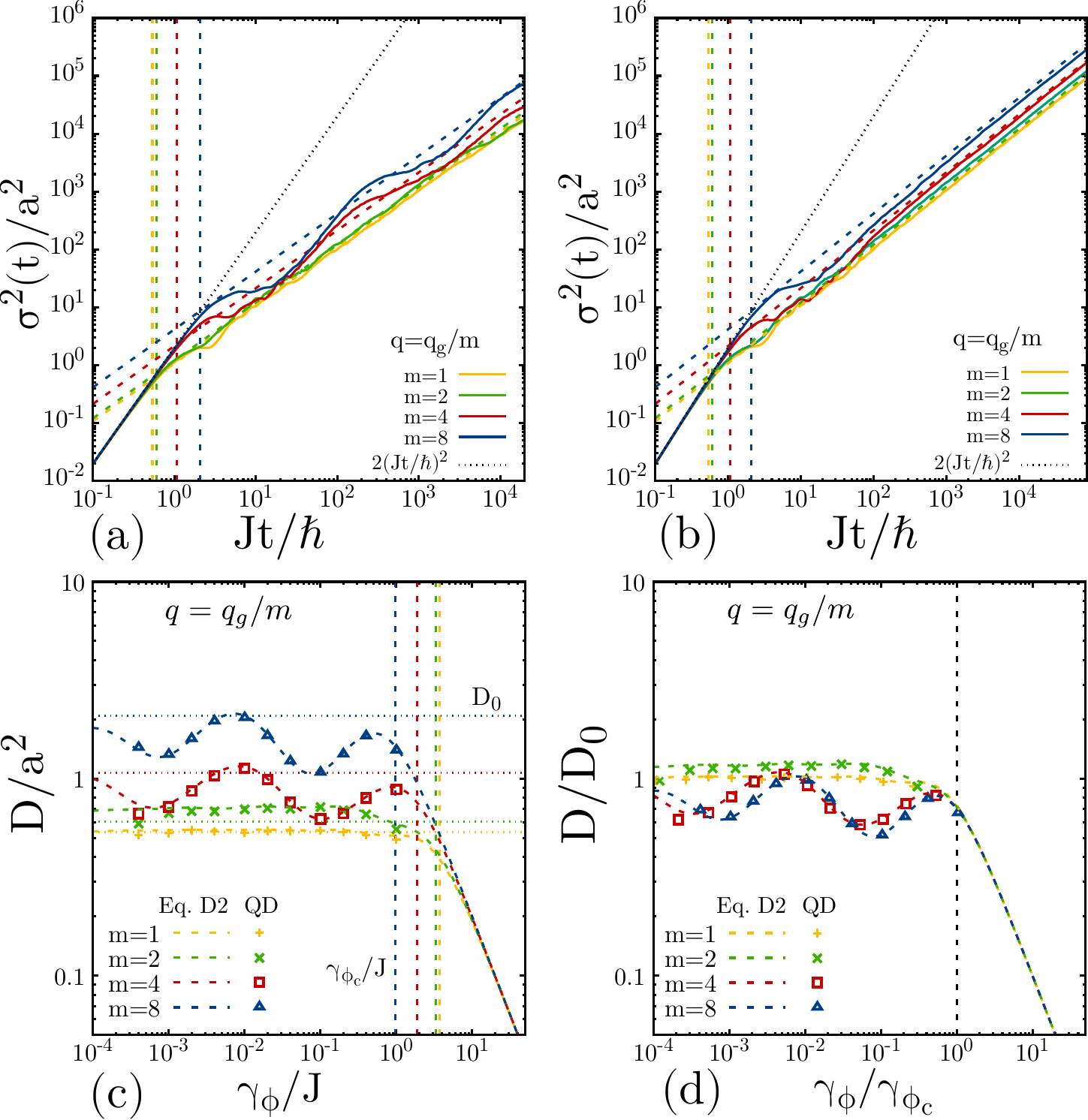}
\caption{Upper panels: Time evolution of the spreading of an excitation in a HHAA chain with $q=q_g/m$ at criticality. The vertical-dashed lines show $\tau_W$ (Eq. \eqref{eq:sup:tauW}), the crosswise dashed lines correspond to $\sigma^2(t)=2D_0t$ (Eq. \eqref{SM:eq:Dt}), while the dotted line shows the initial ballistic evolution (Eq. \eqref{SM:eq:Sig-IniBall}). \textbf{(a)} $\gamma_\phi=0$, \textbf{(b)} $\gamma_\phi = 0.02$. 
Bottom panels: \textbf{(c)}: Diffusion coefficient as a function of the dephasing strength $\gamma_\phi$. $D$ is calculated from the QD dynamics (symbols) and from Eq. \eqref{SM:eq:Jianshu} (dashed lines). The horizontal-dotted shows $D_0$ (Eq. \eqref{SM:eq:Dt}) and the vertical-dashed lines $\gamma^c_\phi=\frac{2\hbar}{\tau_W}$. \textbf{(d)}: The diffusion coefficient and dephasing strength are re-scaled by $D_0$ and $\gamma^c_\phi$ respectively. In the four panels, colors (gray-tones) indicate different values of $m$. Parameters are $N=10000$, $W=2J$.}
\label{fig:Sup:ChQ1}
\end{figure}

The diffusion coefficient derived for the critical point in absence of dephasing (Eq. \eqref{SM:eq:Dt}) shows a dependence on $q$. In order to check the validity of our analytical prediction and the generality of the dephasing-independent regime we analyzed other irrational values of $q$, beyond the golden mean value used in the main text. 

Particularly we study the dynamics of the system using fractions of the golden ratio as irrational numbers $q=q_g/m$, where $m$ is an integer power of two. The continued fractions of the irrationals used are presented in Table I. Trials with irrational numbers of the form $[0,\{m\}]$ yielded similar results.

The spreading in time of the wave packet in absence and presence of dephasing together with our analytical estimations for the diffusion coefficient is shown in Fig.~\ref{fig:Sup:ChQ1}a,b. As one can see, the initial ballistic spreading (Eq. \eqref{SM:eq:Sig-IniBall}) lasts until a time $\tau_W$ (Eq.~(\ref{eq:sup:tauW}), indicated as vertical lines in Fig.~\ref{fig:Sup:ChQ1}a,b). After that time the dynamics is diffusive with a diffusion coefficient given by Eq.~\eqref{SM:eq:Dt}. We notice, see panel (a), the presence of oscillations in the second moment which increase as $q$ decreases. These oscillations are partly erased in presence of dephasing at long times as shown in Fig.~\ref{fig:Sup:ChQ1}b for $\gamma_\phi=0.02$. 

Fig.~\ref{fig:Sup:ChQ1}c shows the fitted values of $D$ (symbols) together with the $D$ values obtained from Eq. \eqref{SM:eq:Jianshu} (dashed curves) as a function of $\gamma_\phi$ for different $q$ at the MIT. As vertical dashed lines, we plot $\gamma^c_\phi=\frac{2\hbar}{\tau_W}$ which coincide with the beginning of the strong dephasing regime, where the diffusion coefficient decreases with dephasing. Notice that for large values of $m$, the diffusion coefficient $D$ exhibits significant oscillations with respect to $\gamma_\phi$. This phenomenon arises from the observed oscillations in the coherent dynamics (Eq. \eqref{SM:eq:DSigma0-gen}), likely due to the weaker irrationality of the $q$ value compared to $q_g$. More investigations should be done in the future to understand the origin of these interesting oscillations. In Fig.~\ref{fig:Sup:ChQ1}d we plot the diffusion coefficient re-scaled by the theoretical value in absence of dephasing (Eq. \eqref{SM:eq:Dt}) and $\gamma_\phi$ rescaled by the elastic scattering rate $\gamma^c_\phi=\frac{2\hbar}{\tau_W}$. Fig.~\ref{fig:Sup:ChQ1}d confirm the validity of our analytical expressions of $D$ and $\tau_W$ as a function of $q$. 

\begin{table}
\caption{Continued fraction of the irrational used in this section $q=q_g/m$. The numbers between brackets are infinitely repeated in the fraction.}
\begin{tabular}{l }
\hline
\hline
 Continued Fraction  \\ 
\hline
$\frac{\sqrt{5}-1}{2}$ = $[0,\{1\}]$  =     $\dfrac{1}{1 +\dfrac{1}{1+\cdots}}$    \\
\\
$\frac{1}{2}\frac{\sqrt{5}-1}{2}$ = $[0,3,\{4\}]$   =   $\dfrac{1}{3 +\dfrac{1}{4+\dfrac{1}{4 + \cdots}}}$ \\ 
\\
$\frac{1}{4}\frac{\sqrt{5}-1}{2}$ = $[0,6,\{2,8\}]$  =  $\dfrac{1}{6 +\dfrac{1}{2+\dfrac{1}{8 + \dfrac{1}{2+\cdots}}}}$   \\ 
\\
\hline
\end{tabular}
\end{table}

\pagebreak
\newpage

\section{Study of different paradigmatic models of transport.}\label{SM:TestModels}

In this section we test the validity of Eq.~\eqref{SM:eq:Gen-tauW} and Eq. \eqref{SM:eq:DSigma0-gen}, using two models that present a coherent diffusion regime and/or criticality: the Fibonacci chain, and the Power-Banded Random Matrices (PBRM) model. 

\subsection{The Fibonacci chain.}\label{SM:FibChain}

\begin{figure}
\centering
\includegraphics[height=6 cm]{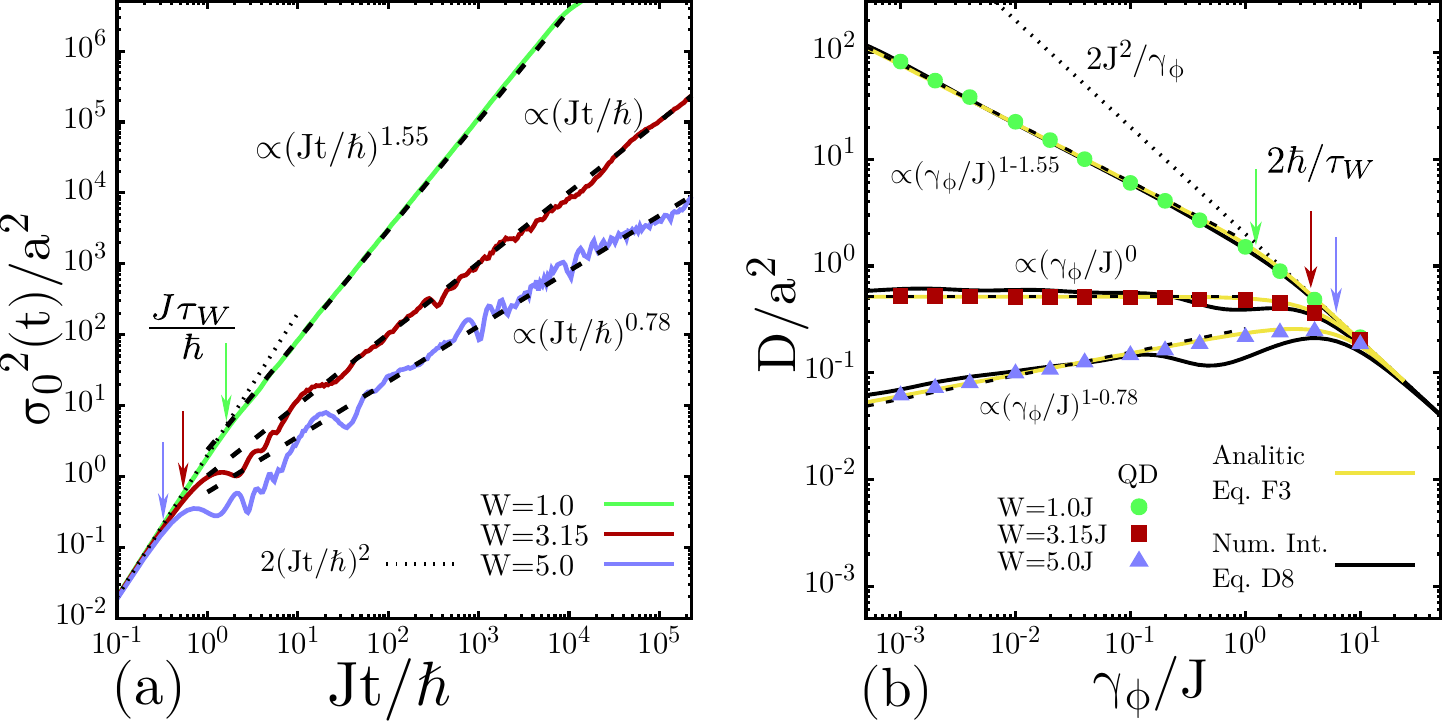}
\caption{\textbf{(a)} Time evolution of the spreading in absence of dephasing for $W=\{1,3.15,5\}$ in the Fibonacci chain (colored (gray-toned) solid curves). The dotted-black lines show the initial ballistic spreading (Eq.~\eqref{SM:eq:Sig-IniBall}), the vertical arrows $\tau_W$ (Eq.~\eqref{SM:eq:Gen-tauW}), and the dashed-black lines show the power law behavior after $\tau_W$. \textbf{(b)} Diffusion coefficient as a function of the dephasing strength. The symbols are fitted directly from the QD dynamic in presence of dephasing, the solid-black curves show Eq.~\eqref{SM:eq:DSigma0-gen} numerically integrated using a Poisson process, and as solid-yellow (light-gray) curves the analytical expression Eq.~\eqref{SM:eq:Fib-D-Analitic} is plotted. The vertical arrows correspond to $2\hbar/\tau_W$, where the strong dephasing regime indicated by a dotted-black line starts. The dashed-black lines indicate the power dependence for small $\gamma_\phi$. The simulations were done in a chain of length $N=10^4$.}
\label{fig:Sup:Fibonacci}
\end{figure}

The Fibonacci model is described by the Hamiltonian:
\begin{equation}
\mathcal{H}=\sum_nJ(\ket{ n}\bra{n+1}+\ket{n+1}\bra{n})
+\varepsilon_n\ket{n}\bra{n},
\end{equation}
where the on-site potential is determined by $\varepsilon_n=W(\lfloor (n+1) q_g^2\rfloor-\lfloor n q_g^2\rfloor)$, here $\lfloor x\rfloor$ represents the integer part of $x$ and $q_g=\frac{\sqrt{5}-1}{2}$ is the golden ratio. In this potential, $\varepsilon_n$ corresponds to the $n$-th element of the Fibonacci word sequence, that can also be obtained by repeated concatenation: $0,0W,0W0,0W00W,0W00W0W0$,etc.

The dynamics in the Fibonacci chain were studied in absence and presence of dephasing \cite{varma2019diffusive,lacerda2021dephasing,ChiPur+Go22}. In absence of dephasing it is known that the second moment grows, after the initial quadratic spreading, as a power law $\sigma^2_0(t) \propto t^{\alpha}$ with an exponent that depends on the strength of the on-site potential. It grows subdiffusively ($\alpha<1$) for $W>3.15J$, diffusively ($\alpha=1$) for $W=3.15J$ and superdiffusively ($\alpha>1$) for $W < 3.15J$. These dynamics are shown in Fig.~\ref{fig:Sup:Fibonacci}a.

This spreading can be written analytically in the approximated and simplified form:
\begin{equation}
\sigma_{0}^{2}(t)=\left\{\begin{array}{ll}
v_{0}^{2} t^{2} & \text { if } t<\tau_{W} \\
2 A t^{\alpha} & \text { if } t>\tau_{W}
\end{array} \text { with } A=\frac{v_{0}^{2} \tau_{W}^{2-\alpha}}{2}.\right. \\
\end{equation}
From this expression and using Eq.~\eqref{SM:eq:DSigma0-gen} with a Poisson process we obtain an analytical expression for the diffusion coefficient in presence of dephasing:
\begin{equation}
D=\frac{v^2_0 \left(\tau_W^3 E_{-\alpha }\left(\frac{\tau_W}{\tau_\phi }\right)+\alpha  \tau_\phi  \tau_W^2 \Gamma (\alpha ) \left(\tau_\phi ^{\alpha } \tau_W^{-\alpha}-\left(\frac{\tau_W}{\tau_\phi }\right)^{-\alpha }\right)+2 \tau_\phi^3-\tau_\phi  e^{-\frac{\tau_W}{\tau_\phi }} \left(2 \tau_\phi ^2+2 \tau_\phi  \tau_W+\tau_W^2\right)\right)}{2 \tau_\phi^2},
\label{SM:eq:Fib-D-Analitic}
\end{equation}
where $\Gamma (\alpha )$ is the Euler Gamma function, and $E_{-\alpha }\left(\frac{\tau_W}{\tau_\phi }\right)=\int^{\infty}_1e^{-\left(\frac{\tau_W}{\tau_\phi }\right)t}t^{\alpha}dt$.

The vertical lines in Fig.~\ref{fig:Sup:Fibonacci}a, represent $\tau_W$, calculated from the analytical equations Eq.~\eqref{eq:Delta} and Eq.~\eqref{SM:eq:Gen-tauW} yielding $1/\tau_W=q_g W/\hbar$. After this time, the initial ballistic dynamic stops and the algebraic dynamic starts. Particularly, for $W=3.15J$, when the subsequent dynamic is diffusive, we obtain $D_0=\frac{v_0^2\tau_W}{2}$. This analytical prediction is shown as a black-dashed line on top of the red (dark-gray) curve.

Once dephasing is added, the dynamics becomes diffusive for all values of $W$. The diffusion coefficient as a function of the dephasing strength was computed numerically through a quantum drift dynamics for different values of $W$. These results are shown as symbols in Fig.~\ref{fig:Sup:Fibonacci}b. They are compared with the numerical integration of Eq.~\eqref{SM:eq:DSigma0-gen} using a Poisson process (black curves) and the analytical expression (Eq. \eqref{fig:Sup:Fibonacci}, yellow (light-gray) curves). We conclude that the diffusion coefficient depends only on the coherent dynamics and the noise strength.

From equations~\eqref{SM:eq:Dsigma0} and~\eqref{SM:eq:Fib-D-Analitic}, it is clear that the dependence of $\sigma_0^2(t)$ determines the behavior of $D(\gamma_\phi)$. Particularly, if $\sigma_0^2(t)\propto t^\alpha$ then $D(\gamma_\phi) \propto (\gamma_\phi)^{(1-\alpha)}$ for $\gamma_\phi\ll 2\hbar/\tau_W$. This dependence is pointed out in Fig.~\ref{fig:Sup:Fibonacci}b with dashed-black lines on top of the data. These results are consistent with recent findings reported in Ref.~\cite{lacerda2021dephasing}.

\subsection{The PBRM model.}

The power-law banded random matrix (PBRM) model describes one-dimensional (1D) tight-binding chains of length $N$ with long-range random hoppings. This model is represented by $N\times N$ real symmetric random matrices whose elements are statistically independent random variables characterized by a normal distribution with zero mean and variance given by, 
\begin{equation}
\braket{|\mathcal{H}_{ii}|^2} =J^2 \text{  and  }	\braket{ |\mathcal{H}_{ij}|^2} = J^2\frac{1}{2} \frac{1}{1+(|i-j|/b)^{2 \mu}}\text{  with  } i\neq j.
	\label{eq:pbrm}
\end{equation}

The PBRM model, Eq.~\eqref{eq:pbrm}, depends on two control parameters: $\mu$ and $b$, while $J$ is an energy scale that can be considered equal to 1 for all practical purposes. 
For $\mu > 1$ ($\mu < 1$) the PBRM model is in the insulating (metallic) phase, so its eigenstates are localized (delocalized). At the MIT, which occurs for all values of $b$ at $\mu = 1$, the eigenfunctions are known to be multi-fractal. 

The statistical properties of the eigenfunctions and eigenvalues of this model have been widely studied\cite{mirlin_transition_1996,mirlin_multifractality_2000,evers_fluctuations_2000,cuevas_anomalously_2001,evers_anderson_2008}. Here we study the spreading dynamics of an initially localized excitation at the middle of the chain in absence and presence of a decoherent environment. 

As in the previous systems, the initial spreading of the local excitation is ballistic, where the second moment is given by $\sigma^2_0=v_0^2t^2$. Generalizing Eq.~(\ref{SM:eq:Sig-IniBall}) to account for the randomness of the Hamiltonian, we found that the velocity $v_0$ is:
\begin{equation}
v^2_0= 2\sum^{N/2}_{n=1} \langle\mathcal{H}_{n,0}^2\rangle n^2= \sum^{N/2}_{n=1} \frac{J^2}{1+(n/b)^{2\mu}} n^2,
\label{SM:PBRM:v0}
\end{equation}
where we summed over the sites to the right and left (factor 2) of the initial site (denoted as 0). This initial velocity (Eq. \eqref{SM:PBRM:v0}) diverges for $\mu < 3/2$ at large $N$ as $N^{3-2\mu}$. For large $N$, $b\ll 1$, and $\mu< 3/2$, the sum can approximated by an integral, yielding $v_0^2\approx J^2 b^{2\mu}\frac{N^{3-2\mu}}{(3-2\mu)2^{3-2\mu}}$. 

This initial ballistic spreading lasts up to $t=\tau_W$, which should be addressed numerically since Eq.~\eqref{eq:Delta} is only valid for NN chains and a similar analysis with this model do not yield a simple expression. However, on a first approximation if we use Eq.~\eqref{eq:Delta}, with uncorrelated and Gaussian distributed site energies with $\braket{|\mathcal{H}_{ii}|^2} =J^2$, we obtain $\tau_W=1$.

\begin{figure}
\centering
\includegraphics[height=6 cm]{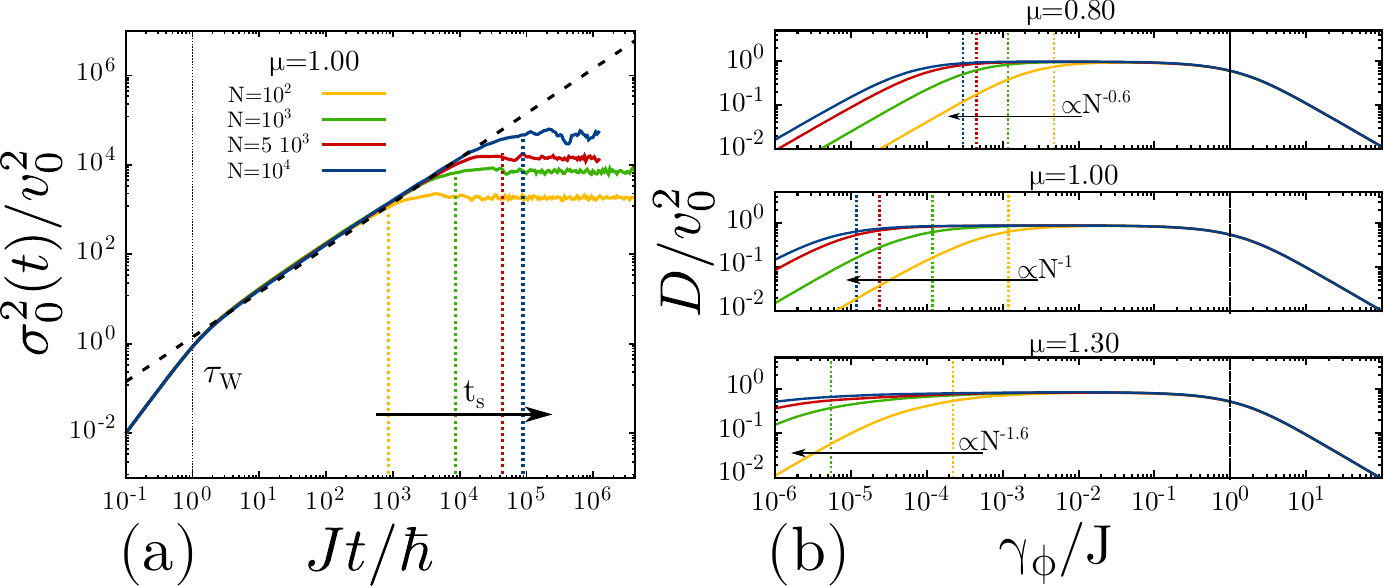}
\caption{\textbf{(a)} Time evolution of the spreading of an initially localized excitation in absence of dephasing in the PBRM model for $\mu=1$, $b=0.01$, and $N=\{100,1000,5000,10000\}$ (colored (gray-toned) curves). The vertical lines denote $\tau_W$ (black) and $t_s$ (colored/gray-tones), while the crosswise dashed line represents the theoretical diffusive spreading $\sigma^2_0(t)/v^2_0\approx \sqrt(2)t$. \textbf{(b)} From top to bottom the figures shows the diffusion coefficient obtained through Eq.~\eqref{SM:eq:Jianshu} for $\mu=\{0.80,1.00,1.30\}$ (colored (gray-toned) curves). The vertical dashed-black line marks the characteristic dephasing where the dynamics start to be dominated by noise and the initial ballistic dynamics (strong Zeno regime). The colored (gray-toned) dashed vertical lines show the values of $\gamma_\phi=2\hbar/t_s$ below which finite size effect starts to be relevant, the dependence with $N$ of the values is indicated in each plot.}
\label{fig:Sup:PBRM-SumFig1}
\end{figure}

For $t>\tau_W$, we find numerically that for $0.5<\mu<1.5$ the second moment of the excitation spreads diffusively (see Fig. \ref{fig:Sup:PBRM-SumFig1}a for $\mu=1$). Note that the parameter $b$ modifies the initial velocity and the diffusion coefficient. Consequently, we choose a small $b=0.01$ to reduce both the magnitude of the initial spread and the diffusion coefficient, generating a slower dynamic and having a larger window for diffusive dynamics before the system reaches saturation (at fix $N$). In the diffusive regime, we find $\sigma^2_0\approx  v_0^2 (\sqrt{2}\tau_W) t$. The factor $\sqrt{2}$ is introduced based on the numerical results to correct the discrepancy in $\tau_W$ due to the long-range hopping. We numerically explore different values of $b$, spanning from $0.001$ to $0.3$, affirming the reliability of our findings (results not presented).

It's important to note that, although the system is localized for $1.0<\mu<1.5$, its eigenfunctions have power-law tails with exponent $2\mu$, therefore its second moment diverge $N\rightarrow\infty$. The presence of these fat tails allow an unbounded growth in time of the the second moment in the limit of $N\rightarrow\infty$. For $\mu<1.5$ the saturation value of the second moment $\sigma^2_{0,SV}$ is $\sigma^2_{0,SV} = \frac{N^2}{12}f(b,\mu)$, where $f(b,\mu)\leq 1$.
 
Thus, for $\mu<1.5$ and assuming a spreading form $\sigma^2_0(t>\tau_W)= v_0^2\tau_W^2+\sqrt{2}v_0^2\tau_W(t-\tau_W)$, we can calculate the time $t_{s}$ where the spreading reaches its saturation value by imposing $\sigma^2_{0,SV}=\sigma^2_0(t_s)$ obtaining:
 \begin{equation}
t_s=\frac{\sigma^2_{0,SV}}{\sqrt{2}v_0^2\tau_W}+\tau_W\frac{(\sqrt{2}-1)}{\sqrt{2}}\propto N^{2\mu-1}.
\label{SS:ts}
\end{equation}
Our analytical estimate of $t_s$ agrees with the numerical finding (see vertical lines in Fig. \ref{fig:Sup:PBRM-SumFig1}a for $\mu=1$). Eq.~\eqref{SS:ts} implies that as $N$ increases, for $\mu<1/2$ the saturation value will be reached at shorter times and eventually the dynamics will be always ballistic ($t_s$ becomes smaller than $\tau_W$). On the opposite case, for $1/2<\mu<3/2$, $t_s$ increases with $N$ and we have a diffusive spreading until saturation. 

As in the previous models, the presence of a coherent quantum diffusion (for $1/2< \mu < 3/2$), generates an almost decoherence-independent diffusive regime. Indeed, for $2\hbar/t_s \lesssim \gamma_\phi \lesssim 2\hbar/\tau_W$, $D$ is almost constant, as most of the {\it environmental measurements} fall in the diffusive regime (after $\tau_W$ and before the saturation time $t_s$). When $\gamma_\phi \ll 2\hbar/t_s$ the noise enters in the dynamics after saturation, generating finite size effects.
 Fig. \ref{fig:Sup:PBRM-SumFig1}b shown $D$ vs. $\gamma_\phi$ computed using the Green-Kubo approach for different $N$ in the extended, critical and localized regimes ($\mu=\{0.8,1,1.3\}$). From Eq.~(\ref{SS:ts}) we can see that for $1/2< \mu < 3/2$, $t_s$ increases with $N$, and finite sizes effects start at smaller values of the decoherence strength, see black arrows in Fig. \ref{fig:Sup:PBRM-SumFig1}b. For $\gamma_\phi > 2\hbar/\tau_W$, decoherence affects the dynamics mainly during the initial ballistic spreading, leading to a decrease of the diffusion coefficient proportional to $v_0^2$.

For $\mu<1/2$, the velocity of the initial ballistic spreading, see Eq.~(\ref{SM:PBRM:v0}), increases with $N$ faster than that saturation value. Therefore, $t_s$ decreases with $N$, becoming smaller than $\tau_W$ and leaving no place for a diffusive dynamic. Hence, no decoherence-independent region can be found for the diffusion coefficient.

For $\mu>3/2$, $t_s$ converges to a constant value as $N$ increases. Thus, for $\gamma_\phi<2\hbar/t_s$ the diffusion coefficient will be linearly dependent on $\gamma_\phi$ and we can not have a dephasing independent regime. This situation is similar to the localized case of the Harper-Hofstadter-Aubry-André.

\section{Purity/Loschmidt Echo.}\label{SM:purity}
\begin{figure}
\centering
\includegraphics[width=1.0\columnwidth]{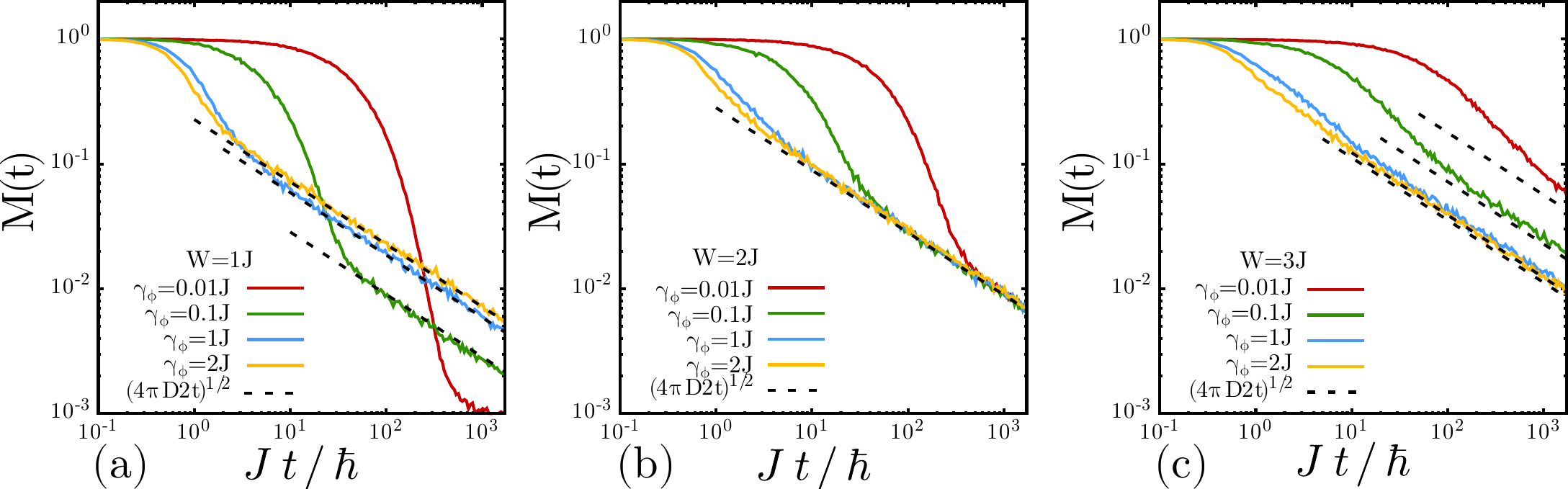}
\caption{Time evolution of the purity (Loschmidt echo) $M(t)$ with different dephasing strength in a HHAA chain with $N=1000$. \textbf{(a)} $W=J$, extended phase. \textbf{(b)} $W=2J$, MIT. \textbf{(c)} $W=3J$, localized phase. Colored (gray-toned) curves represent different $\gamma_\phi$. The dashed-black lines are theoretical predictions ($M(t)\propto\frac{1}{\sqrt{Dt}}$) where $D$ was obtained from Eq. \eqref{SM:eq:Jianshu} for Fig. (a) and (c) and from Eq. \eqref{SM:eq:Dt} for (b).}
\label{fig:SM:purity}
\end{figure}

The purity, defined as,
\begin{equation}
M(t)=\Tr [\rho(t)\rho(t)],
\end{equation}
here $\rho(t)=e^{\mathcal{L}t}\rho_0$ is the evolved density matrix of a local excitation, is a measure of the coherence's level of $\rho(t)$. $M(t)=1$ implies that $\rho(t)$ is a pure state (fully coherent), while $M(t)<1$ indicates a mixed state (incoherent superposition). 

In the following we show that the purity can be calculated using the quantum drift (QD) simulation by generating a Loschmidt echo in the dynamics.

The superoperator $\mathcal{L}$ is defined by,
\begin{equation}
{\cal L}[\rho] = -\frac{i}{\hbar} \left[ \mathcal{H} \rho - \rho \mathcal{H} \right] + {\mathcal L_\phi}[\rho]= {\cal L}_0+{\mathcal L_\phi},
\end{equation}
where $\mathcal{H}$ is the Hamiltonian and ${\mathcal L_\phi}$ the HS dephasing. We can see that $\mathcal{L}^\dagger=\mathcal{L}_0^\dagger+\mathcal{L}_\phi^\dagger=-\mathcal{L}_0+\mathcal{L}_\phi$, and since the density matrix is a Hermitian operator, we have, $\rho(t)=\rho^{\dagger}(t)\implies e^{\mathcal{L}t}\rho_0=\rho_0e^{\mathcal{L}^{\dagger}t}$. Using these properties we rewrite the definition of the purity in the following form,
\begin{eqnarray}
M(t)&=&\Tr [\rho(t)\rho(t)]=\Tr [e^{\mathcal{L}t}\rho_0e^{\mathcal{L}t}\rho_0]\\
&=&\Tr [\rho_0e^{\mathcal{L}^\dagger t} e^{\mathcal{L}t}\rho_0]\equiv \Tr [\rho_0 \rho_{LE}(2t)],
\end{eqnarray}
where it is clear that the purity is a comparison between the initial density matrix and the density matrix $\rho_{LE}(2t)$ which is the result of two evolutions. In detail, there is an initial forward evolution $\rho(t)=e^{(\mathcal{L}_0+\mathcal{L}_\phi)t}\rho_0$ and a second evolution with the sign of the Hamiltonian inverted (backward evolution) $\rho_{LE}(2t)=e^{(-\mathcal{L}_0+\mathcal{L}_\phi)t}\rho(t)$, i.e. the purity corresponds to the echo observed on $\rho_0$ after reverting the time. If the initial state is a pure state $\rho_0=\ket{0}\bra{0}$, we can directly obtain the purity numerically by a stochastic simulation of the forward and backward evolution and by looking at the probability of returning to the initial state (in our case, the initial site).

We studied the purity/Loschmidt echo as a function of time in the extended, critical, and localized regimes in the HHAA model changing the decoherence strength. These results are shown in Figures \ref{fig:SM:purity}. We observe that for short times the decay of the purity is exponential and only depends on the decoherence strength and the initial state in all regimes. After $t\approx4\hbar/\gamma_{\phi}$ (numerically estimated), the decay of the purity becomes a power law, $M(t)\propto\frac{1}{\sqrt{D(\gamma_\phi,W)t}}$, where $D(\gamma_\phi,W)$ is the diffusion coefficient of the forward dynamics (dashed curves in Fig. \ref{fig:SM:purity}). From the results of the previous sections (for $\gamma_\phi<\gamma^c_\phi$) we infer that the rate of decay of the purity in this power-law regime decreases with $\gamma_\phi$ in the extended regime, increases in the localized regime, and remains constant at the critical point. This can be interpreted by considering that the localized states are more protected from decoherence, as decoherence affects fewer sites. In this case, as we increase the decoherence strength the decay of the purity is stronger in both the short and long time regimes as a consequence of the delocalization of the wave function. Secondly, in the extended regime, while a stronger decoherence causes a faster decay in the purity at short times, at large times, where the forward dynamics determine the decay rate, it becomes slower for stronger decoherence. This counter-intuitive result is understood as a consequence of the ballistic growth of the wave packet, which in the large time makes it more sensitive to fluctuations.

\begin{figure}
\centering
\includegraphics[height=6 cm]{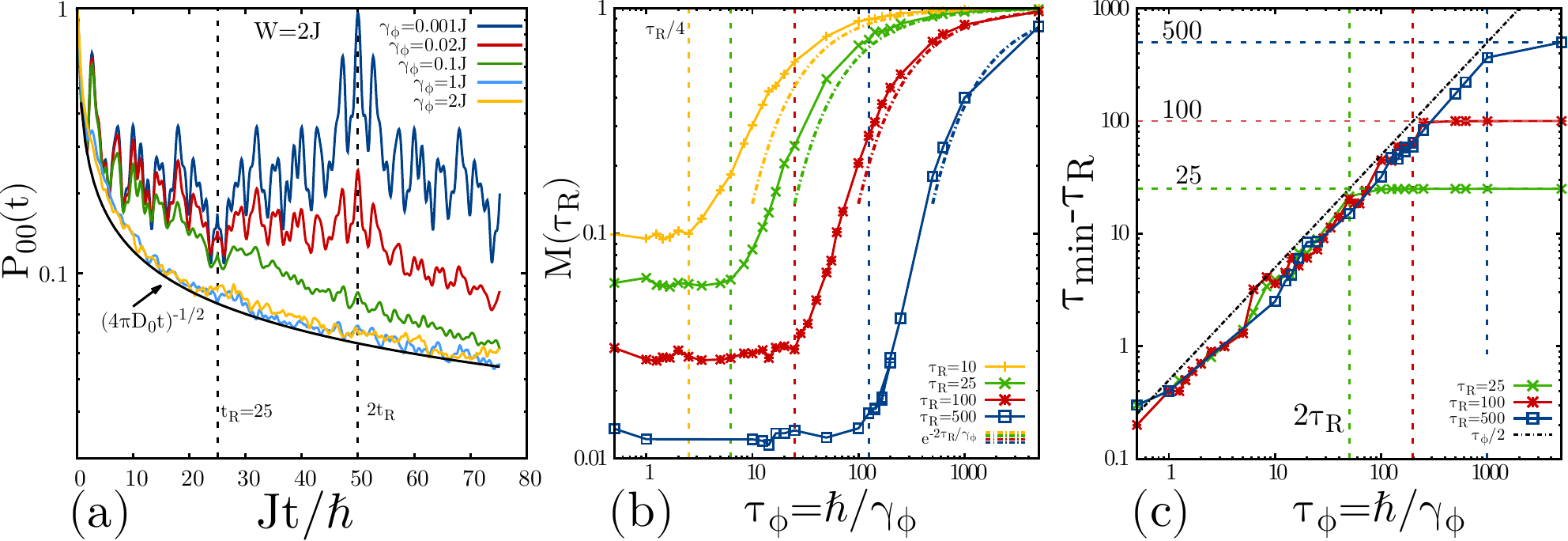}
\caption{\textbf{(a)} Probability to find the particle in the initial site $P_{00}(t)$ for a HHAA chain with $W=2J$. Up to time $\tau_R$ ($\tau_R=25$, first vertical dashed line) the system evolves with $\mathcal{L}$, hereon the sign of the Hamiltonian is inverted (the system evolves with $\mathcal{L}^\dagger$). The value of $P_{00}(t)$ at $2\tau_R$ (echo, second vertical dashed line) corresponds to the purity of the system at the time $\tau_R$ ($M(\tau_R)$). Colored (gray-toned) curves represent different $\gamma_\phi$. \textbf{(b)} Purity (echo) at a fixed time $\tau_R=\{10,25,100,500\}$ as a function of the dephasing time $\tau_\phi=\hbar/\gamma_\phi$ in a HHAA chain with $W=2J$. The vertical colored (gray-toned) dashed lines mark $\tau_R/4$, while the colored (gray-toned) dash-dotted lines show the analytical behavior for $\tau_\phi<<\tau_R$. \textbf{(c)} Time at which the variance of the wave packet reaches its minimum after a Hamiltonian inversion at $\tau_R$ in a HHAA chain with $W=2J$. Vertical colored (gray-toned) dashed lines represent $2 \tau_R$ while the horizontal ones stand for $\tau_R$.}
\label{fig:Sup:EchoPur}
\end{figure}

To clarify the behavior of $M(t)$ at the MIT, we show in Fig. \ref{fig:Sup:EchoPur}a the evolution of $P_{00}$ (probability of being at the initial site), where the Hamiltonian is reverted at time $\tau_R$. At the LE-time, $t=2\tau_R$, one has $P_{00}(2\tau_R)=M(\tau_R)$. For $\gamma_\phi\ll \hbar/\tau_R$, we observe that the $P_{00}(t)$ returns to the initial site and an echo is formed. Note that in absence of dephasing the return is complete and the purity is 1. However, if $\tau_R\gg 4\hbar/\gamma_{\phi}$, $P_{00}(2\tau_R)$ is only determined by the forward diffusive dynamic without a significative echo formation. There are no coherences left to reconstruct the initial dynamic and therefore no echo (peak) is observed, i.e. the $P_{00}(t)$ keeps decaying even with the Hamiltonian reverted. This means that the memory of the initial state is completely lost. Thus, the density matrix is the incoherent superposition of all possible histories. In this sense, after $4\hbar/\gamma_\phi$ the diffusive spreading observed at the MIT differs from the coherent quantum diffusion in the fact that the dynamics it is no longer reversible. 

This purity behavior at the MIT is summarized Fig. \ref{fig:Sup:EchoPur}b, where the value of the echo (purity) for different $\tau_R$ are shown as a function of $\tau_\phi=\hbar/\gamma_\phi$, as one can see, we observed a constant plateau up to $\tau_\phi\approx \tau_R/4$ indicated by vertical dashed-black lines. After that, we have an exponential growth up to the value 1. 

Similar results are found by looking at the width of the returned packet. This is shown in Fig. \ref{fig:Sup:EchoPur}c, where the time at which the second moment reaches its minimum (counted from the reversal time $\tau_R$), is plotted as a function of $\tau_\phi$. After the change in the Hamiltonian sign the wave function starts to shrink, however, this shrinking lasts until the echo time ($2\tau_R$) only if $\tau_\phi>2\tau_R$. This is shown in Fig. \ref{fig:Sup:EchoPur}c as a plateau. When $\tau_\phi<2\tau_R$, the width of the wave packet reaches its minimum at approximately $t\approx\tau_\phi/2$ and starts to broaden again. It is interesting to note that for $ 2\hbar/\tau_R< \gamma_\phi < 4\hbar/\tau_R$, the wave function is widening again but we observe an echo in the polarization. 

We observed that the dependence of the diffusion coefficient with the dephasing strength is inherited by purity (LE) dynamics, as for long times it decays with a power law depending only on $D$. As a consequence, the purity decay at the critical point enters an almost dephasing-independent decay. However, this regime differs substantially from the chaos-induced LE perturbation independent decay proposed by Jalabert $\&$ Pastawski\cite{JaPa01}, as we might have hinted from Ref. \cite{vattay2014quantum}. Indeed, in our case the correlation length of the noise fluctuations is smaller than the mean free path, which does not satisfy the conditions needed for a perturbation-independent decay of the LE. For our local noise, the Feynman history that has suffered a collision with the noisy potential loses the memory of where it comes from, thus it is irreversible as in the Büttiker's dephasing voltage probe. In that sense, the environment-independent decay of the LE/purity, should not be interpreted in the perturbation-independent decoherence context, but rather as a strong irreversibility. 

\end{document}